\documentclass{aa}  

\usepackage{graphicx}
\usepackage{txfonts}
\usepackage{natbib}
\bibpunct{(}{)}{;}{a}{}{,} 
\usepackage{hyperref}
\usepackage{ulem}
\usepackage{color}

\begin{document}

    \titlerunning{Spectropolarimetry of Type II SNe (I)}  
    \authorrunning{T. Nagao et al.}

   \title{Spectropolarimetry of Type~II supernovae}

   \subtitle{(I) Sample, observational data and interstellar polarization
   }

   \author{
            T.~Nagao,\inst{1,2,3}\fnmsep\thanks{takashi.nagao@utu.fi} 
            S.~Mattila,\inst{1,4} 
            R.~Kotak,\inst{1} 
            \and
            H.~Kuncarayakti,\inst{1,5} 
          }

   \institute{
            Department of Physics and Astronomy, University of Turku, FI-20014 Turku, Finland
            \and
            Aalto University Mets\"ahovi Radio Observatory, Mets\"ahovintie 114, 02540 Kylm\"al\"a, Finland
            \and
            Aalto University Department of Electronics and Nanoengineering, P.O. BOX 15500, FI-00076 AALTO, Finland
            \and
            School of Sciences, European University Cyprus, Diogenes street, Engomi, 1516 Nicosia, Cyprus
            \and
            Finnish Centre for Astronomy with ESO (FINCA), FI-20014 University of Turku, Finland
             }

   \date{Received April ??, 2023; accepted ?? ??, 2023}

 
  \abstract
  {We investigate polarization spectra of hydrogen-rich core-collapse supernovae (Type~II SNe). The polarization signal from SNe contains two independent components: intrinsic SN polarization and interstellar polarization (ISP). From these components, we can study the SN explosion geometry and the dust properties in their host galaxies or in the Milky Way. In this first paper, using a new improved method, we investigate the properties of the ISP components of 11 well-observed Type~II SNe. As a result of our analysis, we find that 10 out of these 11 SNe showed a steady ISP component with a polarization degree $\lesssim 1.0$ \%, while one SN was consistent with zero ISP. As for the wavelength dependence, SN~2001dh (and possibly SN~2012aw) showed a non-Milky-Way-like ISP likely originating from the interstellar dust in their respective host galaxies: their polarization maxima were located at short wavelengths ($\lesssim4000$~\AA). Similar results have been obtained previously for highly reddened SNe. The majority of the SNe in our sample had too large uncertainties in the wavelength dependence of their ISP components to consider them further. Our work demonstrates that, by applying this method to a larger SN sample, further investigation of the ISP component of the SN polarization can provide new opportunities to study interstellar dust properties in external galaxies.
  }


   \maketitle

%

\section{Introduction} \label{sec:intro}
The polarization signal from supernovae (SNe) contains information on the explosion geometry, which is directly related to the explosion mechanism. Past studies using polarimetry have revealed that core-collapse SNe are generally aspherical explosions \citep[e.g.,][for a review]{WangWheeler2008}. At the same time, the SN polarization usually also carries an extrinsic component (i.e., interstellar polarization; ISP), which originates from the differential extinction of the SN light by aspherical interstellar (IS) dust grains aligned with the IS magnetic field. It is important to properly remove this ISP component from polarimetric observations of SNe in order to estimate the intrinsic SN polarization, which enables us to study their explosion geometries.

There have been several attempts to estimate the ISP components from the spectropolarimetry of SNe \citep[e.g.,][]{Trammell1993, Tran1997, Wang1997, Leonard2000, Leonard2001, Leonard2002b, Leonard2005, Howell2001, Wang2001, Chornock2006, Patat2012, Mauerhan2015, Inserra2016, Reilly2017, Stevance2017, Stevance2019, Nagao2019, Nagao2021}. These methods rely mainly on either of the following assumptions: (1) the emission peaks of the P-Cygni lines have no intrinsic polarization due to multiple resonant scatterings and/or re-combinations of the photons; (2) the polarization signal from SNe at a sufficiently late phase, when the electron density in the ejecta becomes small enough, is purely from the ISP; (3) a high level of polarization that remains constant in time ($\gtrsim 1$ \%) is mainly from the ISP.

Precise estimation of the ISP component in SN polarization is important not only for deriving the SN intrinsic polarization, but also for investigating the properties of the IS dust. In fact, as SNe are intrinsically bright, using the ISP component in the SN polarization signal leads itself to probing the dust properties in distant galaxies beyond the Milky Way (MW).
It is known that the peak wavelength of the ISP ($\lambda_{\rm{max}}$) is related to the total-to-selective extinction ratio ($R_{V}$), as $R_{V} \sim 5.5\lambda_{\rm{max}}$ $\lbrack \mu$m$\rbrack$ \citep[][]{Serkowski1975}, which is a key parameter of extinction. From the polarization maxima of several reddened SNe at wavelengths shorter \citep[$\lambda_{\rm{max}} \lesssim 0.4$~$\mu$m; \textit{e.g.},][]{Patat2015, Zelaya2017, Chu2022, Nagao2022} than the typical MW ISP  \citep[$\lambda_{\rm{max}} \sim 0.545$~$\mu$m;][]{Serkowski1975}, the presence of peculiarly small dust grains compared to the MW dust has been inferred. At the same time, some reddened Type~II SNe \citep[SN~1999gi and SN~2022ame;][]{Leonard2001, Nagao2022} show MW-like polarization curves for their host-galaxy ISP ($\lambda_{\rm{max}}\sim 0.53$~$\mu$m), implying the existence of MW-like dust in their host galaxies. These facts indicate that there is a diversity in the properties of the IS dust in different galaxies and/or in different locations in galaxies \citep[][]{Nagao2022}.

SNe typically have some intrinsic polarization ($\lesssim 1$ \%)\citep[e.g.,][]{WangWheeler2008}. Therefore, investigations of the properties of IS dust in external galaxies using SN polarization have previously mainly been applied to highly reddened SNe, whose ISP dominates the polarization signal. The application of this method to real observations has been very limited, mainly due to the following reasons: (1) SNe with high extinction are intrinsically rare; (2) such highly reddened objects are optically faint, and therefore rejected for follow-up; (3) spectropolarimetry of faint objects requires access to large telescopes.

In this paper (Paper~I), we describe and implement an improved method compared to those presented in \citet{Nagao2019,Nagao2021} to estimate the ISP component from spectropolarimetric observations of 11 H-rich core-collapse SNe (hereafter Type~II, including Type~IIP and IIL)\footnote{In this work, the classification between Type~IIP and IIL is based on the light-curve shapes of SNe, i.e., the decline rate during the photospheric phase (see a follow-up paper, Paper~II, for details).}, i.e., SNe whose ISP is embedded in their intrinsic SN polarization. 

The paper is organized as follows. In Section~2, we present our SN sample. In Section~3, we present our observational data and describe our data reduction procedures. Then, we explain our method to estimate the ISP and determine and discuss the properties of the ISP in Section~4. At the end, we conclude our findings in Section~5.

In a follow-up paper (Paper~II), we discuss the properties of the intrinsic SN polarization of 15 SNe from our SN sample, their photometric and spectroscopic properties, and the relations between these properties.

\section{SN spectropolarimetric sample}
\label{sec:sample}

We have, from the literature and archives, collected all the publicly-available spectropolarimetric data of Type~II SNe that have multi-epoch observations. 
Our sample consists of 11 Type~II SNe observed with the FOcal Reducer/low-dispersion Spectrograph \citep[FORS1 and FORS2, hereafter FORS;][]{Appenzeller1998} mounted on the Very Large Telescope (VLT) at the European Southern Observatory (ESO), in addition to 4 Type~II SNe observed with other instruments. 
The details of the SNe in our sample are summarised in Table~\ref{tab:sn_basic} and also in the following subsections.

\subsection{SN~2017gmr}
SN~2017gmr was discovered by the `Distance Less Than 40 Mpc' supernova search \citep[DLT40;][]{Tartaglia2018} on 4.27 September 2017 UT \citep[58000.27 MJD;][]{Valenti2017}, in the host galaxy NGC~0988 ($z=0.005037\pm0.000017$; HIPASS Catalog, \citealt{Koribalski2004} via NED\footnote{NASA/IPAC Extragalactic Database.}). It was classified as a core-collapse SN within a few days after the discovery \citep[][]{Pursimo2017}. 
The last non-detection of the object was on 2.23 September 2017 UT (57998.23 MJD), i.e. about two days before the discovery \citep[][]{Valenti2017}. We adopt 57999.09 MJD as the explosion date of SN~2017gmr, distance modulus of $\mu=31.46\pm0.15$ mag, and a total line of sight extinction of $A_{V}=1.14$ mag \citep[][]{Andrews2019}.
%

\subsection{SN~2017ahn}
SN~2017ahn was discovered on 8.29 February 2017 UT \citep[57792.29 MJD;][]{Tartaglia2017}, by DLT40.
It is located in NGC~3318 at $z=0.009255\pm0.000021$ (HIPASS Catalog, via NED), and was classified as a Type~II SN within a day after the discovery \citep[][]{Hosseinzadeh2017}. 
The last non-detection was on 7.23 February 2017 UT (57791.23 MJD), i.e., about one day before the discovery \citep[][]{Tartaglia2017}. We adopt 57791.76 MJD as the explosion date of SN~2017ahn \citep[][]{Tartaglia2021}, $\mu=32.59\pm0.43$ mag  \citep[][]{Sorce2014}, and $A_{V}=0.82$ mag \citep{Tartaglia2021}.
%

\subsection{SN~2013ej}
SN~2013ej was discovered by the Lick Observatory Supernova Search on 25.45 July 2013 UT \citep[56498.45 MJD;][]{Kim2013}. 
It is located in Messier~74 at $z=0.002192\pm0.000003$ \citep[][]{Lu1993}. 
The last non-detection of the object was on 23.54 July 2013 UT (56496.54 MJD), i.e. about two days before the discovery \citep[][]{Shappee2013}. We adopt 56496.9 MJD as the explosion date of SN~2013ej \citep[][]{Dhungana2016}, $\mu=29.91\pm0.16$ mag \citep[][]{Huang2015}, and $A_{V}=0.19$ mag \citep[][]{Dhungana2016}.
%

\subsection{SN~2012ec}
SN~2012ec was discovered on 11.04 August 2012 UT \citep[56150.04 MJD;][]{Monard2012}. 
It is located in NGC~1084 at $z=0.004693\pm0.000013$ (from the HIPASS Catalog via NED). The object was classified as a Type~II SN on 17.7 August 2012 UT \citep[][]{Takaki2012}. 
The last non-detection of the object was on 28.75 March 2012 UT \citep[56014.75 MJD;][]{Monard2012}. We adopt 56143.00 MJD as the explosion date of SN~2012ec \citep[][]{Barbarino2015}, $\mu=31.10\pm0.06$ mag \citep[][]{Springob2009},  and $A_{V}=0.45$ mag \citep[][]{Barbarino2015}.
%

\subsection{SN~2012dh}
SN~2012dh was discovered on 26.98 June 2012 UT \citep[56104.98 MJD;][]{Maza2012}. 
It is located in ESO~443-G80 at $z=0.007055\pm0.000007$ (from the HIPASS Catalog via NED). The object was classified as a Type~II SN by the spectra obtained on 28.84 June, 30.84 June and 2.84 July 2012 UT \citep[][]{Maza2012}.
The last non-detection of the object was on 23.11 June 2012 UT \citep[56101.11 MJD;][]{Maza2012}. Based on the spectrum obtained on 7 July 2012 UT with the Nordic Optical Telescope, \citet[][]{Fraser2012} suggested that this SN was discovered soon after explosion. We adopt the discovery date of 56104.98 MJD as the explosion date of SN~2012dh and $\mu=31.85\pm0.45$ mag \citep[][]{Tully2016}. From the lack of significant interstellar Na~I~D absorption lines originated from the host galaxy in our spectra, we ignore the extinction in the host galaxy. We only take the extinction value of $A_{V}=0.20$ mag from the MW extinction into account \citep[][]{Schlafly2011}.
%

\subsection{SN~2012aw}
SN~2012aw was discovered on 16.86 March 2012 UT \citep[56002.86 MJD;][]{Fagotti2012}.
It is located in M~95 at $z=0.002595\pm0.000013$ (from the HIPASS Catalog via NED). The object was classified as a Type~II SN on 19.5 March 2012 UT \citep[][]{Itoh2012}. 
The last non-detection of the object was on 15.27 March 2012 UT \citep[56001.27 MJD;][]{Poznanski2012}. We adopt 56002.10 MJD as the explosion date of SN~2012aw \citep[][]{Barbarino2015}, $\mu=30.01\pm0.09$ mag \citep[][]{deJaeger2019}, and $A_{V}=0.23$ mag \citep[][]{Bose2013}.
%

\subsection{SN~2010hv}
SN~2010hv was discovered on 18.18 August 2010 UT \citep[55426.18 MJD;][]{Pignata2010}.
It is located in 2MASX~J23242536-1902139 at $z=0.024280\pm0.000043$ (from the HIPASS Catalog via NED). 
%
The last non-detection of the object was on 2.15 August 2010 UT \citep[55410.15 MJD;][]{Pignata2010}. We adopt $55418.2 \pm 8.0$ MJD, which is the middle of the discovery and last non-detection date, as the explosion date of SN~2010hv and consider the time from the last non-detection to the discovery as its uncertainty.
%

\subsection{SN~2010co}
SN~2010co was discovered on 6.105 May 2010 UT \citep[55322.105 MJD;][]{Monard2010}.
It is located in NGC~6862 at $z=0.014026\pm0.000017$ (from the HIPASS Catalog via NED). The object was classified as a Type~II SN on 20 May 2010 UT \citep[][]{Morrell2010, Green2010}.
The last non-detection of the object was on 13.144 April 2010 UT \citep[55299.144 MJD;][]{Monard2010}. We adopt $55310.6 \pm 11.5$ MJD, which is the middle of the discovery and last non-detection date, as the explosion date of SN~2010co and consider the time from the last non-detection to the discovery as its uncertainty.
%

\subsection{SN~2008bk}
SN~2008bk was discovered on 25.141 March 2008 UT \citep[54550.141 MJD;][]{Monard2008}.
It is located in NGC~7793 at $z=0.000767\pm0.000013$ (from the HIPASS Catalog via NED). The object was classified as a Type~II SN on 12.4 April 2008 UT \citep[][]{Morrell2008}. 
The last non-detection of the object was on 2.742 January 2008 UT \citep[54467.742 MJD;][]{Monard2008}. We adopt 54549.50 MJD as the explosion date of SN~2008bk \citep[][]{VanDyk2012a}, $\mu=27.66\pm0.04$ mag \citep[][]{Zgirski2017}, and $A_{V}=0.07$ mag \citep[][]{VanDyk2012a}.
%

\subsection{SN~2007aa}
SN~2007aa was discovered on 18.308 February 2007 UT \citep[54149.308 MJD;][]{Doi2007}.
It is located in NGC~4030 at $z=0.004887\pm0.000013$ (from the HIPASS Catalog via NED). The object was classified as a Type~II SN on 19.24 February 2007 UT \citep[][]{Folatelli2007}. 
The last non-detection of the object was on 12.36 March 2005 UT \citep[][]{Doi2007}. We adopt 54126.7 MJD as the explosion date of SN~2007aa, which was estimated in \citet[][]{Gutierrez2017}.
We adopt the distance modulus of $\mu=32.38\pm0.20$ mag towards SN~2007aa, as estimated in \citet[][]{Tully2013}. From the lack of significant interstellar Na~I~D absorption lines originated from the host galaxy in our spectra and the minimal interstellar polarization \citet[][]{Chornock2010}, we ignore the extinction in the host galaxy. We only take the extinction from the MW, $A_{V}=0.07$ mag, into account \citet[][]{Schlafly2011}.
%

\subsection{SN~2006ov}
SN~2006ov was discovered on 24.86 November 2006 UT \citep[54063.86 MJD;][]{Nakano2006}.
It is located in NGC~4303 at $z=0.005224\pm0.000007$ (from the HIPASS Catalog via NED). The object was classified as a Type~II SN on 25.56 November 2006 UT \citep[][]{Blondin2006}. 
The last non-detection of the object was on 4 May 2006 UT \citep[][]{Nakano2006}. We adopt 53974 MJD as the explosion date of SN~2006ov \citep[][]{Li2007} and $\mu=30.5\pm0.4$ mag \citep[][]{Spiro2014}. From the lack of significant interstellar Na~I absorption lines originated from the host galaxy in our spectra and the minimal interstellar polarization \citep[][]{Chornock2010}, we ignore the extinction in the host galaxy. We only take the MW extinction, $A_{V}=0.06$ mag, into account \citep[][]{Schlafly2011}.


\subsection{SN~2004dj}
SN~2004dj was discovered on 31.76 July 2004 UT \citep[53217.76 MJD;][]{Nakano2004}.
It is located in NGC~2403 at $z=0.000445\pm0.000001$ (from the HIPASS Catalog via NED). The object was classified as a Type~II SN on 3.17 August 2004 UT \citep[][]{Patat2004}. 
The last non-detection of the object was on 11 October 2002 UT \citep[][]{Nakano2004}. We adopt 53186.5 MJD as the explosion date of SN~2004dj \citep[][]{Chugai2005}, $\mu=27.70\pm0.19$ mag, and $A_{V}=0.22$ mag \citep[][]{Vinko2006}.
%

\subsection{SN~2001du}
SN~2001du was discovered on 24.7 August 2001 UT \citep[52145.7 MJD;][]{Evans2001}.
It is located in NGC~1365 at $z=0.005457\pm0.000003$ (from the HIPASS Catalog via NED). The object was classified as a Type~II SN on 2.05 September 2001 UT \citep[][]{Smartt2001}.
The last non-detection of the object was on 23.7 August 2001 UT \citep[52144.7 MJD;][]{Evans2001}. We adopt $52145.2 \pm 0.5$ MJD, which is the middle of the discovery and last non-detection date, as the explosion date of SN~2001du and consider the term from the last non-detection to the discovery as its uncertainty.
%

\subsection{SN~2001dh}
SN~2001dh was discovered on 22.88 July 2001 UT \citep[52112.88 MJD;][]{Chassagne2001}.
It is located in ESO~340-G9 at $z=0.008700\pm0.000017$ (from the HIPASS Catalog via NED). The object was classified as a Type~II SN on 8 August 2001 UT \citep[][]{Patat2001}.
The last non-detection of the object was on 26.1 June 2001 UT \citep[52086.1 MJD;][]{Chassagne2001}. We adopt $52099.5 \pm 13.4$ MJD, which is the middle of the discovery and last non-detection date, as the explosion date of SN~2001dh and consider the term from the last non-detection to the discovery as its uncertainty.
%

\subsection{SN~1999em}
SN~1999em was discovered on 29.44 October 1999 UT \citep[51480.44 MJD;][]{Li1999}.
It is located in ESO~340-G9 at $z=0.008700\pm0.000017$ (from the HIPASS Catalog via NED). The object was classified as a Type~II SN on 29.7 October 1999 UT \citep[][]{Deng1999}.
The last non-detection of the object was on 20.45 October 1999 UT \citep[51471.45 MJD;][]{Li1999}. We adopt 51475.1 MJD as the explosion date of SN~1999em \citep[][]{Leonard2002a}, $\mu=30.34\pm0.24$ mag \citep[][]{Leonard2003}, and $A_{V}=0.31$ mag \citep[][]{Leonard2002a}.
%

\begin{table*}
\caption{Details of the SNe}
\label{tab:sn_basic}
$
  \begin{tabular}{lllllll} \hline
     & Type & Host galaxy & t$_{\rm{exp}}$ (MJD) & Redshift & $\mu$ (mag) & $A_{V}$ (mag) \\ 
    \hline\hline
    SN~2017gmr & IIP & NGC~988 & 57999.09 & 0.005037 & 31.46 $\pm$ 0.15 & 1.14\\ \hline
    SN~2017ahn & IIL & NGC~3318 & 57791.76 & 0.009255 & 32.59 $\pm$ 0.43 & 0.82\\ \hline
    SN~2013ej & IIL & M~74 & 56496.90 & 0.002192 & 29.91 $\pm$ 0.16 & 0.19\\ \hline
    SN~2012ec & IIP & NGC~1084 & 56143.00 & 0.004693 & 31.10 $\pm$ 0.06 & 0.45\\ \hline
    SN~2012dh & IIL & ESO~443-G80 & 56104.98 & 0.007055 & 31.85 $\pm$ 0.45 & 0.20\\ \hline
    SN~2012aw & IIP & M~95 & 56002.10 & 0.002595 & 30.01 $\pm$ 0.09 & 0.23\\ \hline
    SN~2010hv & II & 2MASX~J23242536-1902139 & 55418.20 $\pm$ 8.00 & 0.024280 & - & - \\ \hline
    SN~2010co & II & NGC6862 & 55310.60 $\pm$ 11.50 & 0.014026 & - & - \\ \hline
    SN~2008bk & IIP & NGC~7793 & 54549.50 & 0.000767 & 27.66 $\pm$ 0.04 & 0.07\\ \hline
    SN~2007aa & IIP & NGC~4030 & 54126.70 & 0.004887 & 32.38 $\pm$ 0.20 & 0.06\\ \hline 
    SN~2006ov & IIP & NGC~4303 & 53974.00 & 0.005224 & 30.5 $\pm$ 0.4 & 0.06\\ \hline 
    SN~2004dj & IIP & NGC~2403 & 53186.50 & 0.000445 & 27.70 $\pm$ 0.19 & 0.22\\ \hline 
    SN~2001du & II & NGC~1365 & 52145.20 $\pm$ 0.50 & 0.005457 & - & - \\ \hline
    SN~2001dh & II & ESO~340-G9 & 52099.50 $\pm$ 13.40 & 0.008700 & - & - \\ \hline
    SN~1999em & IIP & NGC~1637 & 51475.10 & 0.002392 & 30.34 $\pm$ 0.24 & 0.31\\ \hline
  \end{tabular}
  $
  
  \begin{minipage}{\hsize}
        \smallskip
        Notes. The SN type is based on the light-curve shape (see Paper~II). $t_{\rm{exp}}$ is the estimated explosion date (see Section~\ref{sec:sample}). The redshifts for the SNe are assumed to be those of their host galaxies, given by NASA/IPAC Extragalactic Database (NED). The references for the distance module are given in Section~\ref{sec:sample}. $A_{V}$ is the Galactic extinction in the V band.
    \end{minipage}
\end{table*}

\section{Observational data and data reduction}

\subsection{Spectropolarimetric data}
We have collected spectropolarimetric data of the VLT sample through the ESO Science Archive Facility\footnote{\url{http://archive.eso.org}}, reanalyzed the data, and derived the values for the ISP and the intrinsic SN polarization. The details of the spectropolarimetric observations of the VLT sample are shown in Table~\ref{tab:sn_sample1}. The observation logs are given in Appendix~\ref{app:obs_log}. 
For increasing the signal-to-noise ratio, we combined the specstropolarimetric data obtained at similar epochs checking the consistency of the polarization signal.
We note that the data qualities of the observations at the last (third) epoch for SN~2010co are too bad to obtain the polarization signal, where the polarization due to the photon shot noise and the variations within different observation frames in this epoch are larger than the polarization signal. Therefore, we have not used the last epoch of the data for SN~2010co in our analysis.
For the SNe observed with other instruments, we have instead adopted the derived values for the ISP and the intrinsic SN polarization from the literature: \citet[][]{Leonard2001} for SN~1999em, \citet[][]{Leonard2006} for SN~2004dj and \citet[][]{Chornock2010} for SN~2006ov and SN~2007aa.

For the VLT sample, the details of the instrumentation and the data reduction procedure are the same as outlined in \citet[][]{Nagao2019}. All observations used the optimal set of half-wave retarder-plate (HWP) angles, \textit{i.e.}, 0, 45.0, 22.5 and 67.5 degrees \citep[e.g.,][]{Patat2006} and employed the low-resolution G300V grism. The observations were taken with different CCDs, and some observations did and others did not use an order-sorting filter (see Table~\ref{tab:sn_sample1}). The raw data were reduced using standard methods described in \citet[][]{Patat2006} and IRAF \citep[][]{Tody1986,Tody1993}. The ordinary and extraordinary beams of the spectropolarimetric data were extracted with a fixed aperture size of $10$ pixels. The extracted spectra were resampled to 5 {\AA} bins in Paper~I and 50 {\AA} bins in Paper~II for better signal-to-noise ratios. The HWP zeropoint angle chromatism was corrected based on the tabular data in the FORS2 user manual \footnote{\url{http://www.eso.org/sci/facilities/paranal/instruments/fors/doc/VLT-MAN-ESO-13100-1543_P07.pdf}}. The wavelength scale was corrected to the rest frame based on the host-galaxy redshifts (see Table~\ref{tab:sn_basic}). The polarization bias in the polarimetric spectra was removed according to the standard procedure developed by \citet[][]{Wang1997}.

\begin{table*}
\caption{Details of the SNe in the VLT sample. The filters here are order-sorting filters, where the numbers in the names identify the cut-on wavelength in nm. Where applicable, references to previous publications using the data are provided.
}
\label{tab:sn_sample1}
$
  \begin{tabular}{llllll} \hline
     & Telescope & Instrument & CCD & Filter & Reference \\ \hline\hline
    SN~2017gmr & VLT-UT1 & FORS2 & MIT & -- & \citet[][]{Nagao2019}\\ \hline
    SN~2017ahn & VLT-UT1 & FORS2 & MIT & -- & \citet[][]{Nagao2021}\\ \hline
    SN~2013ej & VLT-UT1  & FORS2 & MIT & GG435 & \citet[][]{Nagao2021}, \citet[][]{Leonard2021}\\ \hline
    SN~2012ec & VLT-UT1 & FORS2 & MIT & GG435 & this work\\ \hline
    SN~2012dh & VLT-UT1 & FORS2 & MIT & GG435 & this work\\ \hline
    SN~2012aw & VLT-UT1 & FORS2 & MIT & GG435 & \citet[][]{Dessart2021}\\ \hline
    SN~2010hv & VLT-UT1 & FORS2 & MIT & GG435 & this work\\ \hline
    SN~2010co & VLT-UT1 & FORS2 & MIT & GG435 & this work\\ \hline
    SN~2008bk & VLT-UT2 & FORS1 & EEV & GG375 & \citet[][]{Leonard2012}\\ \hline
    SN~2001du & VLT-UT3 & FORS1 & Tek & GG435 & this work\\ \hline
    SN~2001dh & VLT-UT3 & FORS1 & Tek & GG435 & this work\\ \hline
  \end{tabular}
  $
\end{table*}

\subsection{Photometric data}

We also collated publicly-available $V$-band light curves for 10 SNe in our sample from the literature (see Table~\ref{tab:photometric_data}). Here, the data of SN~2012ec, SN~2012aw, SN~2008bk and SN~1999em were obtained through the Open Supernova Catalog \citep[][]{Guillochon2017}. For SN~2007aa, we use previously unpublished data from the Calar Alto observatory (Appendix~\ref{app:07aa}). We use Vega-based magnitudes throughout.

\begin{table*}
\caption{References of the light-curve data}
\label{tab:photometric_data}
$
  \begin{tabular}{ll} \hline
    Name & Data source\\ \hline\hline
    SN~2017gmr & \citet[][]{Andrews2019}\\ \hline
    SN~2017ahn & \citet[][]{Tartaglia2021}\\ \hline
    SN~2013ej & Berkeley SuperNova DataBase \citep[SNDB;][]{Silverman2012}\\ \hline
    SN~2012ec &  \citet[][]{Smartt2015}\\ \hline
    SN~2012dh &  Ayala et al. in prep.\\ \hline
    SN~2012aw & \citet[][]{Munari2013, Bose2013, DallOra2014, Brown2014}\\ \hline
    SN~2008bk & \citet[][]{VanDyk2012a, Anderson2014}\\ \hline
    SN~2007aa &  This paper\\ \hline 
    SN~2006ov &  \citet[][]{Spiro2014}\\ \hline 
    SN~2004dj &  \citet[][]{Tsvetkov2008}\\ \hline 
    SN~1999em &  \citet[][]{Anderson2014, Galbany2016}\\ \hline
  \end{tabular}
  $
\end{table*}

\subsection{Spectroscopic data}
We use the spectra obtained as the spectropolarimetric observations for the VLT sample (see Table~\ref{tab:sn_sample1}). These spectra are provided in Appendix~\ref{app:spec}. The spectra of SN~2006ov are taken from \citet[][]{Spiro2014} through the Supernova Open Catalog. In addition, we use our unpublished new spectroscopic data for SN~2007aa (see Appendix~\ref{app:07aa} for details). For SN~2004dj and SN~1999em, we adopt the derived values for the Fe~II velocity in \citet[][]{Takats2012} instead of analysing the spectra by ourselves (see Paper~II).

\section{Interstellar polarization}
\label{sec:ISP}

In this section, we describe our method to determine the ISP component of the SN polarization and apply this method to the spectropolarimetric data of the VLT sample (see Table~\ref{tab:sn_sample1}).

\begin{figure*}
  \includegraphics[width=2\columnwidth]{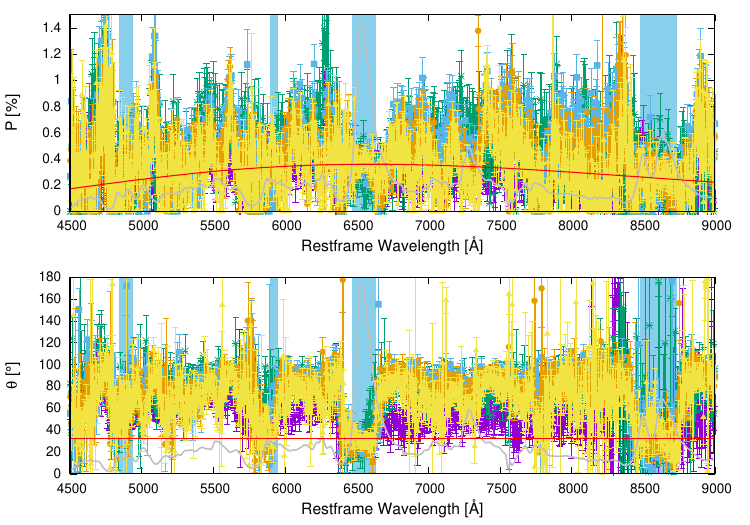}
    \caption{
    Polarization degree $P$ (top panel) and angle $\theta$ (bottom panel) for SN~2017gmr at five different epochs: 58045.60, 58067.24, 58099.86, 58108.76 and 58134.93 (MJD; purple, green, sky-blue, orange and yellow symbols, respectively). The data are binned to $5$ {\AA}. The gray lines at the bottom of each plot are the total-flux spectra on 58134.93 (MJD). The red lines show the polarization degree and angle of the ISP as approximated by $P(\lambda) = P_{\rm{max}} \exp \left[ -K \ln^{2} \left( \lambda_{\rm{max}}/\lambda \right) \right]$ with fixed values of $P_{\rm{max}}=0.35$, $\lambda_{\rm{max}}=6400$ \AA\; and $K=4.0$, and $\theta_{\rm{ISP}}=33.37$. The blue shading shows the wavelength windows at the adopted lines, where at all epochs the flux was $> 1.1$ times higher than in the nearby continuum regions.
    }
    \label{fig:17gmr}
\end{figure*}

\begin{figure*}
  \includegraphics[width=2\columnwidth]{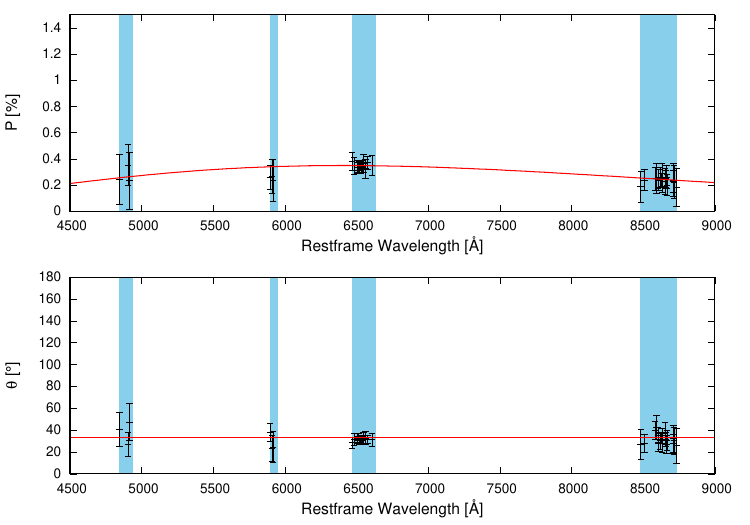}
    \caption{
    Polarization degree $P$ (top panel) and angle $\theta$ (bottom panel) selected for the determination of the ISP in SN~2017gmr. The red lines and the blue-shaded areas are the same as in Fig.~\ref{fig:17gmr}.  In each wavelength bin, the ISP values are the average over all five epochs, after removal of outliers as described in Sect.~\ref{subsec:estimation}. The red lines show the best-fit ISP in Sect.~\ref{subsec:estimation} and Table~\ref{tab:best_fit}.
    }
    \label{fig:ISP_17gmr}
\end{figure*}

\subsection{ISP estimation}
\label{subsec:estimation}
The continuum regions in the spectra of most Type~II SNe are intrinsically polarized due to scattering in an aspherical photosphere. This polarization is time dependent and typically most prominent at the beginning of the tail phase \citep[$\lesssim1.0$\%; \textit{e.g.},][]{WangWheeler2008,Nagao2019}. For the determination of the ISP in Type~II SNe, this fact must be taken into account. For the extraction of the ISP signal from the polarimetric spectra, we adopted a similar strategy as described in \citet[][]{Nagao2019,Nagao2021}, which assumes that emission peaks of strong spectral lines carry no intrinsic polarization signal, due to the depolarization of the underlying continuum by the unpolarized line emission. In reality, the prominent emission lines (\textit{e.g.}, H$\alpha$ $\lambda 6563$ and the Ca II triplet $\lambda 8498/8542/8662$) do display a non-vanishing polarization with an angle that slightly differs from that measured in other spectral regions (see Fig.~\ref{fig:17gmr}). Ten out of the 11 SNe in our sample show such a non-zero polarization at line peaks (see Appendix~\ref{app:pol_spec}), which we attribute to the ISP. Only in SN~2010hv is the polarization at the emission peaks consistent with zero within the uncertainties (see Fig.~\ref{fig:app_10hv}). Hence we assume that its ISP component is zero. All polarization spectra of our sample are shown in Appendix~\ref{app:pol_spec}.

For the ISP measurements, we adopted an improved method of \citet[][]{Nagao2019, Nagao2021}. In this paper, we adopted the strong lines due to H$\beta$ $\lambda 4861$, Fe II $\lambda 4924$, Na~I~D $\lambda 5890/5896$, H$\alpha$ $\lambda 6563$, and the Ca II triplet $\lambda 8498/8542/8662$. We evaluated the ISP in windows where the line flux exceeds 1.1 times the continuum level in the flux spectrum normalized by the continuum (see Appendix~\ref{app:spec}). This assures that the selected signal is dominated by the depolarized line emissions compared to the possibly polarized continuum radiation. We have chosen such ISP regions individually for each epoch in the 5-\AA-binned polarization spectra (blue-shaded regions in Figs.~\ref{fig:17gmr}, \ref{fig:ISP_17gmr}, and \ref{fig:app_17gmr}-\ref{fig:app_01dh}). In the following processes, when calculating an average value using the 5-\AA-binned polarization signals in these ISP regions, we take an average of polarization degrees or angles taking their errors (but not the flux levels) into account. The efficiency of depolarization by line scattering is different for different lines and also depends on the SN phase at the time of the observation. Therefore, assuming that the ISP is time independent, we averaged, in each wavelength bin (5-\AA-bin), the polarization degrees from all epochs. In this process, we removed polarization signals whose polarization degree differed from the average value by more than one standard deviation (1~$\sigma$), in each 5-\AA-bin, and repeated to revise the epoch-averaged value until all used values satisfied the 1~$\sigma$ criteria. Then, assuming that the polarization angle of the ISP has a constant value in wavelengths, we eliminated epoch-averaged polarization signals with polarization angles differing by more than 1~$\sigma$ from the average for all the epoch-averaged polarization signals. After this elimination, we recalculate the average value of the polarization angle for all the remaining epoch-averaged polarization signals, and repeated this elimination process until all remaining values satisfied the 1~$\sigma$ criteria. Finally, for each line, we also excluded polarization signals with degrees deviating by more than 1~$\sigma$ from the averaged value in the line-specific wavelength window, until all surviving values satisfied the 1~$\sigma$ criteria.

As an example, Fig.~\ref{fig:ISP_17gmr} shows the polarization degree and angle of the signals selected as ISP in the polarization spectrum of SN~2017gmr. The average polarization angle of the ISP signals ($\sim32\degr$) clearly differs from that of the continuum component \citep[$\sim 95\degr$;][see also Fig.~\ref{fig:17gmr}]{Nagao2019}. Since the depolarization by line scattering in the relatively weaker lines (H$\beta$, Fe II and Na) is inefficient, the number of selected polarization signals is smaller than for the H$\alpha$ and Ca~II lines.

\subsection{ISP properties}
After the selection of the ISP signals, we calculated the wavelength dependence of the ISP by $\chi^{2}$ fitting of the selected ISP signals with the classical Serkowski function \citep[][]{Serkowski1975}: $P(\lambda) = P_{\rm{max}} \exp \left[ -K \ln^{2} \left( \lambda_{\rm{max}}/\lambda \right) \right]$. We investigated the parameter ranges of $0.0 \leq P_{\rm{max}} \leq 3.0$ [\%], $100 \leq \lambda_{\rm{max}} \leq 10000$ [\AA] and $0.0 \leq K \leq 5.0$ with steps of 0.01 [\%], 100 [\AA] and 0.1, respectively. We calculated the values of the reduced chi-square ($\chi^{2}_{\nu}$) for all possible combinations of the fitting parameters, and obtained the best-fit values and the 1-$\sigma$ confidence levels in a 3-D grid. The best-fit values and their $\chi^{2}_{\nu}$ values are shown in Table~\ref{tab:best_fit}. The values of $\chi^{2}_{\nu}$ ($\lesssim 1$) are satisfactorily small, as expected if the ISP components are well fitted by the \citeauthor{Serkowski1975} function. However, these good fits can also be due to the small numbers of 2-5 wavelength windows used.

\begin{table}
      \caption[]{Best-fit values for the ISP wavelength dependency (see Sect.~\ref{subsec:estimation}).}
         \label{tab:best_fit}
    $\displaystyle
         \begin{array}{lccccc}
            \hline
            \noalign{\smallskip}
            \rm{SN} & \theta_{\rm{ISP}} & P_{\rm{max}} & \lambda_{\rm{max}} & K & \chi^{2}_{\nu} \\
             & (\rm{degree}) & (\%) & ({\AA}) &  & \\
            \noalign{\smallskip}
            \hline\hline
            \noalign{\smallskip}      
            \rm{SN 2017gmr} & 31.9 \pm 0.9 & 0.34 & 6500 & 4.3 & 0.11 \\
            \noalign{\smallskip} \hline \noalign{\smallskip}
            \rm{SN 2017ahn} & 116.2 \pm 1.6 & 0.65 & 6200 & 3.5 & 0.12 \\
            \noalign{\smallskip} \hline \noalign{\smallskip}
            \rm{SN 2013ej} & 105.1 \pm 0.6 & 2.84 & 800 & 0.3 & 0.07 \\
            \noalign{\smallskip} \hline \noalign{\smallskip}
            \rm{SN 2012ec} & 11.1 \pm 3.5 & 0.14 & 10000 & 0.1 & 0.13 \\
            \noalign{\smallskip} \hline \noalign{\smallskip}
            \rm{SN 2012dh} & 53.3 \pm 1.8 & 1.36 & 200 & 0.1 & 0.16 \\
            \noalign{\smallskip} \hline \noalign{\smallskip}
            \rm{SN 2012aw} & 124.6 \pm 0.8 & 2.25 & 500 & 0.2 & 0.47 \\
            \noalign{\smallskip} \hline \noalign{\smallskip}
            \rm{SN 2010hv} & - & 0.0 & - & - & - \\
            \noalign{\smallskip} \hline \noalign{\smallskip}
            \rm{SN 2010co} & 171.1 \pm 3.1 & 1.18 & 6900 & 1.5 & 0.07 \\
            \noalign{\smallskip} \hline \noalign{\smallskip}
            \rm{SN 2008bk} & 164.0 \pm 2.7 & 0.34 & 900 & 0.1 & 0.43 \\
            \noalign{\smallskip} \hline \noalign{\smallskip}
            \rm{SN 2001du} & 11.7 \pm 3.5 & 0.23 & 10000 & 1.2 & 0.08 \\
            \noalign{\smallskip} \hline \noalign{\smallskip}
            \rm{SN 2001dh} & 171.3 \pm 2.2 & 1.39 & 3900 & 4.9 & 0.14 \\
            \noalign{\smallskip}
            \hline
         \end{array}
         $
\end{table}

\begin{figure*}
  \includegraphics[width=2.0\columnwidth]{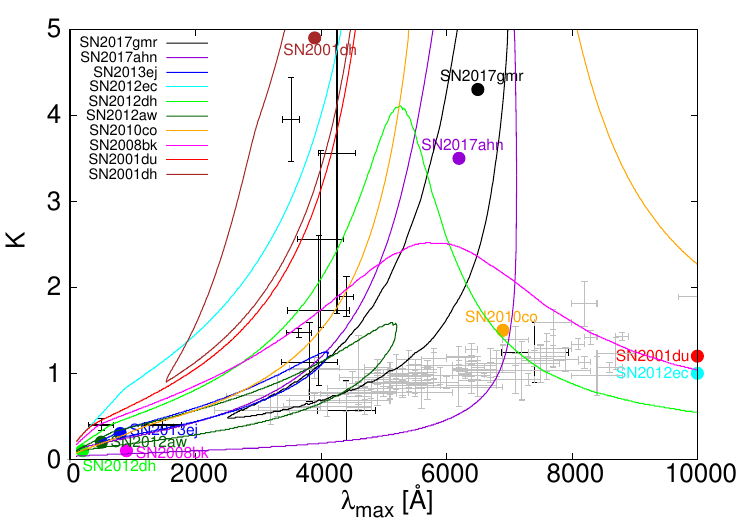}
    \caption{
    The ISP $\lambda_{\rm{max}}$-$K$ diagram for the Type~II SNe in this study.  Several Type~Ia SNe \citep[black crosses;][]{Patat2015,Zelaya2017,Cikota2018} and a large number of MW stars \citep[gray crosses;][]{Whittet1992} are also included. The colored points show the best-fit values of the Type~II SNe, and the lines represent the 1-$\sigma$ confidence intervals for the fitting. 
    }
    \label{fig:ISP1}
\end{figure*}

\begin{figure*}
  \includegraphics[width=2.0\columnwidth]{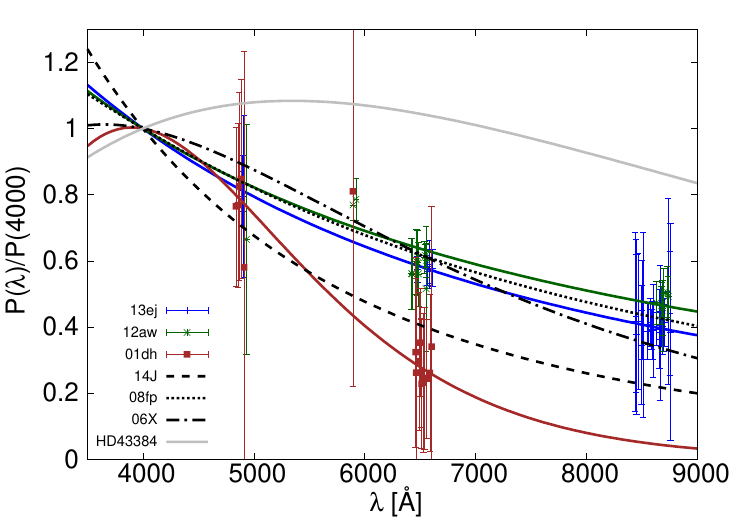}
    \caption{
    Wavelength dependence of the ISP (normalized to the polarization at 4000\AA) towards SN~2013ej, SN~2012aw and SN~2001dh (colored points) with their Serkowski law best fits (colored lines). For comparison, the data of three Type~Ia SNe \citep[SN~2014J, SN~2008fp and SN~2006X;][]{Patat2015} and a Galactic star \citep[HD~43384;][]{Cikota2018} are also plotted.
    }
    \label{fig:ISP2}
\end{figure*}

In Fig.~\ref{fig:ISP1}, we present the best-fit values with their 1-$\sigma$ confidence levels on the $\lambda_{\rm{max}}$-$K$ plane. For some SNe, the confidence areas are too large to allow us to infer similarities with the ISP properties of MW stars or Type~Ia SNe (SN~2017gmr, SN~2017ahn, SN~2012ec, SN~2012dh, SN~2010co, SN~2008bk and SN~2001du). Nevertheless, some objects (SN~2013ej, SN~2012aw and SN~2001dh) do show relatively small confidence areas that are closer to the Type~Ia SN population rather than the MW-star population, even though the areas for SN~2013ej and SN~2012aw include small parts of the MW distribution. In addition, the sample displays large spreads in the values of $\lambda_{\rm{max}}$ and $K$. For example, SN~2001dh deviates from all other SNe, and there is only little overlap between the 1-$\sigma$ confidence areas for SN~2013ej and SN~2017gmr. Such diversity is also seen in the sample of Type~Ia SNe (black crosses in Fig.~\ref{fig:ISP1}). The different observational errors and wavelength windows used in the ISP determination for our sample affect the accuracy of the derived best-fit values for the fitting parameters ($P_{\rm{max}}$, $\lambda_{\rm{max}}$ and $K$) and the fitting errors. However, since the $\chi^{2}_{\nu}$ values are already small for the best fits (see \ref{tab:best_fit}), the current fitting-error areas in Fig.~\ref{fig:ISP1} are the largest for our observational data and method. In the $\lambda_{\rm{max}}$ vs.\ $K$ diagram, the error zones will shrink with an increase of the number of usable wavelength windows and with a reduction of the observational uncertainties. Therefore, with spectropolarimetry of highly extincted SNe whose ISP dominates their polarization, a better determination of the wavelength dependence of the ISP is possible \citep[e.g.,][]{Patat2015,Nagao2022}.

The wavelength dependence of the polarization for the three Type~II SNe with relatively small error ranges (SN~2013ej, SN~2012aw and SN~2001dh) is presented in Figure~\ref{fig:ISP2}. Like for the Type~Ia SNe also included in this figure, their polarization maxima are at wavelengths of 4000~\AA\; or shorter. This is in stark contrast to the typical Galactic star HD43384, the polarization of which peaks near 5500 \AA, although SN~2013ej and SN~2012aw can be consistent with some MW stars within the diversity of MW stars. Some of the fitted $\lambda_{\rm{max}}$ values for Type~II SNe fall outside the wavelength range ($\sim4000$ to $\sim9000$ \AA) of the observations (see Fig.~\ref{fig:ISP1}). However this does not affect the qualitative conclusion of a fundamental difference (mainly in $K$) between SN~2001dh and the MW stars because the ISP peaks for the MW objects are in this wavelength range ($\sim4000$ to $\sim9000$ \AA).

Several of our datasets have been analyzed before (see references in Table~\ref{tab:sn_sample1}). Here, we briefly compare our results with the past ones and discuss the differences. The best-fit ISP derived in our previous papers for SN~2017gmr and SN~2017ahn \citep[][]{Nagao2019,Nagao2021} are consistent within the uncertainties with those presented here (see Figure~\ref{fig:ISP1}): $P_{\rm{max}}=0.4$\%, $\lambda_{\rm{max}}=4900$ \AA\; and $K=1.1$ for SN~2017gmr \citep[][]{Nagao2019} and $P_{\rm{max}}=2.05$\%, $\lambda_{\rm{max}}=200$ \AA\; and $K=0.1$ for SN~2017ahn \citep[][]{Nagao2021}. The values of $\lambda_{\rm{max}}$ and $K$ for SN~2013ej ($P_{\rm{max}}=1.32$\%, $\lambda_{\rm{max}}=3300$ \AA\; and $K=1.0$) derived in \citet[][]{Nagao2021} are only marginally consistent with those in this paper, being slightly outside the 1-$\sigma$ error range. This is because of the mildly modified method applied in the present work. Also, in the previous paper, we used all the emission lines, including weaker lines, which might include incompletely-depolarized signals, while in this paper we limit the fit to the strong lines (H$\beta$, Fe II, Na, H$\alpha$, and the Ca II triplet). In addition, in our previous work, we determined the polarization angle of the ISP as $\theta=110$ degrees by eye, but derived a different value $\theta=105 \pm 0.6$ degrees in this work. These small differences likely give rise to the slightly different best-fit values. 

The ISP estimated by \citet[][]{Leonard2021} for SN~2013ej ($P_{\rm{max}}=0.79$, $\lambda_{\rm{max}}=5100$ \AA\; and $K=1.28$) is slightly different from our values. The reason is that \citeauthor{Leonard2021} only used the H$\alpha$ and Ca II lines and limited the admitted parameter ranges to $0.6 < K < 1.5$ and 5000 \AA\; $< \lambda_{\rm{max}} <$ 5500 \AA, respectively. This leads to a slightly different ISP wavelength dependence from ours, especially at shorter wavelengths. However, the ISP with the largest $\lambda_{\rm{max}}$ in our the 1-$\sigma$ area ($P_{\rm{max}}=1.01$ \%, $\lambda_{\rm{max}}=4100$ \AA\; and $K=1.2$) is reasonably similar, because we used the same data with similar methods. In addition, \citet[][]{Leonard2021} concluded that the ISP of SN~2013ej originated from the MW dust based on the Na~I~D features and that the polarization peak of the ISP should be located around 5250 {\AA} based on the wavelengths of peak polarization observed for two Galactic stars close to the line of sight of SN~2013ej. Taking into account this, the correct ISP for SN~2013ej should be relatively close to the above case on the 1-$\sigma$ border ($P_{\rm{max}}=1.01$ \%, $\lambda_{\rm{max}}=4100$ \AA\; and $K=1.2$).

\citet[][]{Dessart2021} estimated the ISP wavelength dependence for SN~2012aw under slightly different assumptions from ours. Their derived best-fit values ($P_{\rm{max}}=0.45$, $\lambda_{\rm{max}}=5535$ \AA, $K=3.37$ and $\theta=127\degr$) are outside of the 1-$\sigma$ error region of our best-fit values (see Fig.~\ref{fig:ISP1}). This difference probably originates mainly from the different estimates of the polarization degree at blue regions, \textit{e.g.}, around the H$\beta$ line. We notice that the two methods differ in several aspects: \citet[][]{Dessart2021} assumed a relation between $K$ and $\lambda_{\rm{max}}$ derived for the ISP in the MW \citep[$K=1.13 + 0.000405 (\lambda_{\rm{max}}-5500)$;][]{Cikota2018}, while we did not use such a constraint in the fittings of the wavelength dependence of the polarization. On the other hand, we introduced several rejection criteria for incompletely-depolarized signals (see Sect.~\ref{subsec:estimation}). Our estimated polarization degree at H$\beta$ is $\sim0.8$\%, while \citeauthor{Dessart2021} found $\sim 0.4$\%. This difference may be explained by the fact that we used the line-emission regions only, while \citeauthor{Dessart2021} may have included parts of the absorption components. However, the ISP with the largest $\lambda_{\rm{max}}$ in our the 1-$\sigma$ area ($P_{\rm{max}}=0.66$ \%, $\lambda_{\rm{max}}=5200$ \AA\; and $K=1.5$) is reasonably close to their value, because we used the same data with similar methods.

\begin{figure*}
  \includegraphics[width=2.0\columnwidth]{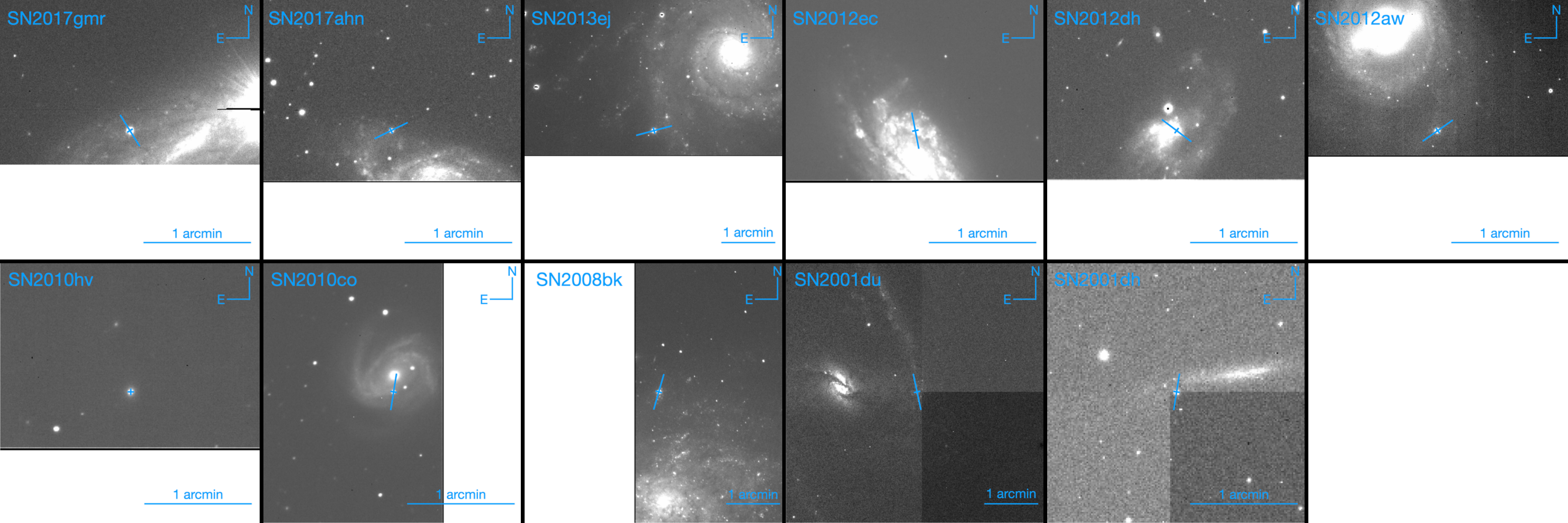}
    \caption{
    VLT/FORS acquisition images of the SNe and their host galaxies (see Table~\ref{tab:sn_basic}). The light blue bars show the polarization angles of the ISP components.
    }
    \label{fig:host}
\end{figure*}

\subsection{ISP due to MW dust vs.\ dust in the host galaxies}
\label{subsec:origin}

ISP can originate not only from dust in the host galaxies, but also from MW dust. The extinction values adopted for the MW reddening of the SNe in our sample are given in Table~\ref{tab:extinction}. From them and the empirical relation between the extinction and the polarization \citep[$P \leq 9E(B-V)$;][]{Serkowski1975}, we can estimate the maximum expected polarization degree due to MW dust ($P_{\rm{max,MW}}$; see Table~\ref{tab:extinction}). Here, this maximum polarization is attained in the extreme case when all the MW-dust grains along the line of sight are aligned in the same direction. Therefore, we expect the MW ISP to fall into the inequality domain of \citeauthor{Serkowski1975}'s relation.

Since for some SNe in our sample (SN~2017ahn, SN~2012ec, SN~2012dh, SN~2010hv and SN~2001du; see Table~\ref{tab:extinction}) the polarization degree at 6000 \AA\; ($P(6000)$; see in Table~\ref{tab:extinction}) is less than (and in some cases comparable to) the maximum values of the \citeauthor{Serkowski1975} relation, their ISP may be partly, or even completely, due to MW dust. This might be the reason why the ISP components of these Type~II SNe are consistent not only with those of Type~Ia SNe, but also with those of MW stars. However, the other Type~II SNe (SN~2017gmr, SN~2013ej, SN~2012aw, SN~2010co, SN~2008bk and SN~2001dh) show an ISP in excess of the \citeauthor{Serkowski1975} maximum, which may indicate that the contribution by host-galaxy dust is significant or even dominates. In fact, the cases of SN~2017gmr and SN~2012aw, the extinction in the host galaxies is estimated to be larger than the MW extinction \citep[see][]{Andrews2019,VanDyk2012b}, and the maximum polarization from the host extinction based on the \citeauthor{Serkowski1975} relation is larger than the observed ISP. On the contrary, since no significant interstellar Na~I~D absorption lines originated from the host galaxy was seen in the high-resolution spectra of SN~2013ej and SN~2008bk \citep[][]{VanDyk2012a, Bose2015}, the above estimation on the large host contribution to the ISP should be wrong for these SNe. It is more reasonable to consider as the origin of the ISP of these SNe is the MW dust and they are extreme cases in the \citeauthor{Serkowski1975} relation, where the polarization degrees are slightly higher than the maximum polarization based on the \citeauthor{Serkowski1975} relation. As for SN~2010co and SN~2001dh, there is no available high-resolution spectrum to quantitatively discuss the extinction in their host galaxies and to check the validity of the above interpretation as the large host contribution to the ISP. The MW dust may contribute to the ISP as significantly as the host-galaxy dust. The MW contribution to the total net polarization degree would be maximal if all the dust grains along the line of sight were aligned in the same direction in both the MW and the host galaxy. Since this is unlikely to be generally the case, we conclude that the ISP of SN~2010co and SN~2001dh likely originates from the dust in their host galaxies.

The ISP angle traces the direction of the magnetic field in the region where the ISP is formed, since this occurs through the differential absorption of the electromagnetic wave by aspherical dust grains aligned with the local magnetic field \citep[\textit{e.g.},][]{Davis1951}. Since the direction of the magnetic field in a spiral galaxy globally follows the direction of the spiral arms \citep[\textit{e.g.},][]{Beck2015}, even though the magnetic field and thus the ISP might suffer local perturbations, \textit{e.g.}, from supernovae \citep{Ntormousi2018}. Type~II SNe generally occur in late-type galaxies, and thus a comparison between the direction of the local spiral arm and the ISP vector may reveal relevant information. Figure~\ref{fig:host} shows the derived ISP angles and VLT/FORS acquisition images of the host galaxies. The ISP pseudo-vectors for SN~2017gmr, SN~2013ej, SN~2012ec, SN~2012dh and SN~2010co are not in any obvious way aligned with the local spiral arm of their hosts, while some correlation is at least plausible in the SN~2017ahn, SN~2012aw, SN~2008bk, and SN~2001du. Since the host of SN~2001dh is viewed close to edge-on, it is difficult to estimate the direction of any structure near the SN location.

The above findings can be summarised as follows: the ISP of SN~2013ej and SN~2008bk should be dominated by MW components. The ISP of SN~2012ec and SN~2012dh might be dominated by MW components. The ISP of SN~2017gmr and SN~2010co might be affected by the MW component. The ISP of SN~2017ahn, SN~2012aw, SN~2001du and SN~2001dh likely originates mainly within their respective host galaxies.

\begin{table}
      \caption[]{Extinction within the Milky Way towards the sample SNe}
      \label{tab:extinction}
    $\displaystyle
         \begin{array}{lccc}
            \hline
            \noalign{\smallskip}
            \rm{SN} & E(B-V)_{\rm{MW}} & P_{\rm{max,MW}} & P(6000)\\
             &  & (\%) & (\%)\\
            \noalign{\smallskip}
            \hline\hline
            \noalign{\smallskip}      
            \rm{SN 2017gmr} & 0.024 & 0.216 & 0.33\\
            \noalign{\smallskip} \hline \noalign{\smallskip}
            \rm{SN 2017ahn} & 0.069 & 0.621 & 0.65\\
            \noalign{\smallskip} \hline \noalign{\smallskip}
            \rm{SN 2013ej} & 0.062 & 0.558 & 0.84\\
            \noalign{\smallskip} \hline \noalign{\smallskip}
            \rm{SN 2012ec} & 0.023 & 0.207 & 0.14\\
            \noalign{\smallskip} \hline \noalign{\smallskip}
            \rm{SN 2012dh} & 0.064 & 0.576 & 0.43\\
            \noalign{\smallskip} \hline \noalign{\smallskip}
            \rm{SN 2012aw} & 0.024 & 0.216 & 0.65\\
            \noalign{\smallskip} \hline \noalign{\smallskip}
            \rm{SN 2010hv} & 0.026 & 0.234 & 0.0\\
            \noalign{\smallskip} \hline \noalign{\smallskip}
            \rm{SN 2010co} & 0.053 & 0.477 & 1.15\\
            \noalign{\smallskip} \hline \noalign{\smallskip}
            \rm{SN 2008bk} & 0.017 & 0.153 & 0.24\\
            \noalign{\smallskip} \hline \noalign{\smallskip}
            \rm{SN 2001du} & 0.018 & 0.162 & 0.17\\
            \noalign{\smallskip} \hline \noalign{\smallskip}
            \rm{SN 2001dh} & 0.052 & 0.468 & 0.56\\
            \noalign{\smallskip}
            \hline
         \end{array}
         $
         \begin{minipage}{.88\hsize}
        \smallskip
        Notes. The values of the extinction were taken from  \citet[][]{Schlafly2011} through NED (NASA/IPAC Extragalactic Database; \url{https://ned.ipac.caltech.edu}). The maximum values of the interstellar polarization were calculated using the empirical relation, $P \leq 9 E(B-V)$, by \citet[][]{Serkowski1975}. The polarization degree at 6000 \AA\; in the best-fit ISP (see Table~\ref{tab:best_fit}) is shown as $P(6000)$.
        \end{minipage}
\end{table}

\section{Conclusions}
\label{sec:conc}
In this paper, we have introduced a method to study the ISP properties using 11 Type~II SNe with a marginal ISP in contrast with the previously studied objects with high extinctions and ISPs. One SN (SN~2010hv) shows zero ISP, while steady ISP components with $\lesssim 1.0$ \% polarization degrees were observed in ten SNe in the sample. As for the wavelength dependence, SN~2001dh (and possibly also SN~2012aw) showed polarization maxima at short wavelengths ($\lesssim 4000$~\AA), which is a similar property as has been previously seen in several highly reddened SNe. The non-MW-like ISP in SN~2001dh and SN~2012aw likely originates from the interstellar dust in their host galaxies.
The ISP in SN~2013ej showed a polarization maximum at a shorter wavelength than the average value of the MW ISP, but probably still within its diversity. This ISP likely originates from the interstellar dust in the MW. 
In the remaining seven SNe (SN~2017gmr, SN~2017ahn, SN~2012ec, SN~2012dh, SN~2010co, SN~2008bk and SN~2001du), the uncertainties of the wavelength dependence are too large to determine whether the ISP is more similar to that seen in the MW or that observed in some reddened SNe.

Previously, the investigation of the IS dust in external galaxies using the SN polarization has been limited mainly to highly reddened SNe, whose ISP dominates their polarization. The IS dust properties of only a few type~II SNe have been investigated \citep[SN~1999gi, SN~2022aau, and SN~2022ame;][]{Leonard2001, Nagao2022}. Our study increased the current SN sample available for investigating the IS dust in the host galaxies of Type~II SNe, demonstrating that further investigation of a larger SN sample can provide new opportunities to study interstellar dust properties in external galaxies.
In particular, spectropolarimetry of SNe whose MW extinction is lower will be better suited to investigate the wavelength dependence of the ISP (i.e., the properties of the interstellar dust) in the SN host galaxies. As a future work, it will be important to study relations between the ISP properties and the global and local properties of the host galaxies, in order to investigate the origin of the diverse properties of the IS dust in external galaxies.

\begin{acknowledgements}
    We thank the anonymous referee for the comments, which substantially improved this paper.
    We are grateful to Stefan Taubenberger for the photometric and spectroscopic data of SN~2007aa and useful comments on the manuscript. We also thank Ferdinando~Patat, Dietrich~Baade, Aleksandar Cikota, Santiago Gonz{\'a}lez-Gait{\'a}n, Keiichi Maeda, Masaomi Tanaka and Mattia~Bulla for useful discussions. 
    This work is based on observations collected at the European Organisation for Astronomical Research in the Southern Hemisphere (ESO) under programmes 099.D-0543, 91.D-0401(A), 089.D-0515(A), 085.D-0391, 081.D-0128(A), 082.D-0151(A) and 67.D-0517(A). This research has made use of the services of the ESO Science Archive Facility. 
    This work is based partly on observations collected at the Centro Astron{\'o}mico Hispano en Andaluc{\'i}a (CAHA) at Calar Alto, operated jointly by Junta de Andaluc{\'i}a and Consejo Superior de Investigaciones Cient{\'i}ficas (IAA-CSIC), under program F07\_2.2\_15.
    T.N. was funded by the Academy of Finland project 328898.
    H.K.\ was funded by the Academy of Finland projects 324504 and 328898.
    S.M. was funded by the Academy of Finland project 350458.
    This research has made use of NASA's Astrophysics Data System Bibliographic Services.
\end{acknowledgements}

%
%

\bibliographystyle{aa} 
\bibliography{aa.bib}

\begin{thebibliography}{106}
\expandafter\ifx\csname natexlab\endcsname\relax\def\natexlab#1{#1}\fi

\bibitem[{{Anderson} {et~al.}(2014){Anderson}, {Gonz{\'a}lez-Gait{\'a}n},
  {Hamuy}, {Guti{\'e}rrez}, {Stritzinger}, {Olivares E.}, {Phillips},
  {Schulze}, {Antezana}, {Bolt}, {Campillay}, {Castell{\'o}n}, {Contreras}, {de
  Jaeger}, {Folatelli}, {F{\"o}rster}, {Freedman}, {Gonz{\'a}lez}, {Hsiao},
  {Krzemi{\'n}ski}, {Krisciunas}, {Maza}, {McCarthy}, {Morrell}, {Persson},
  {Roth}, {Salgado}, {Suntzeff}, \& {Thomas-Osip}}]{Anderson2014}
{Anderson}, J.~P., {Gonz{\'a}lez-Gait{\'a}n}, S., {Hamuy}, M., {et~al.} 2014,
  \apj, 786, 67

\bibitem[{{Andrews} {et~al.}(2019){Andrews}, {Sand}, {Valenti}, {Smith},
  {Dastidar}, {Sahu}, {Misra}, {Singh}, {Hiramatsu}, {Brown}, {Hosseinzadeh},
  {Wyatt}, {Vinko}, {Anupama}, {Arcavi}, {Ashall}, {Benetti}, {Berton},
  {Bostroem}, {Bulla}, {Burke}, {Chen}, {Chomiuk}, {Cikota}, {Congiu}, {Cseh},
  {Davis}, {Elias-Rosa}, {Faran}, {Fraser}, {Galbany}, {Gall}, {Gal-Yam},
  {Gangopadhyay}, {Gromadzki}, {Haislip}, {Howell}, {Hsiao}, {Inserra},
  {Kankare}, {Kuncarayakti}, {Kouprianov}, {Kumar}, {Li}, {Lin}, {Maguire},
  {Mazzali}, {McCully}, {Milne}, {Mo}, {Morrell}, {Nicholl}, {Ochner},
  {Olivares}, {Pastorello}, {Patat}, {Phillips}, {Pignata}, {Prentice},
  {Reguitti}, {Reichart}, {Rodr{\'\i}guez}, {Rui}, {Sanwal}, {S{\'a}rneczky},
  {Shahbandeh}, {Singh}, {Smartt}, {Strader}, {Stritzinger}, {Szak{\'a}ts},
  {Tartaglia}, {Wang}, {Wang}, {Wang}, {Wheeler}, {Xiang}, {Yaron}, {Young}, \&
  {Zhang}}]{Andrews2019}
{Andrews}, J.~E., {Sand}, D.~J., {Valenti}, S., {et~al.} 2019, \apj, 885, 43

\bibitem[{{Appenzeller} {et~al.}(1998){Appenzeller}, {Fricke}, {F{\"u}rtig},
  {G{\"a}ssler}, {H{\"a}fner}, {Harke}, {Hess}, {Hummel}, {J{\"u}rgens},
  {Kudritzki}, {Mantel}, {Meisl}, {Muschielok}, {Nicklas}, {Rupprecht},
  {Seifert}, {Stahl}, {Szeifert}, \& {Tarantik}}]{Appenzeller1998}
{Appenzeller}, I., {Fricke}, K., {F{\"u}rtig}, W., {et~al.} 1998, The
  Messenger, 94, 1

\bibitem[{{Barbarino} {et~al.}(2015){Barbarino}, {Dall'Ora}, {Botticella},
  {Della Valle}, {Zampieri}, {Maund}, {Pumo}, {Jerkstrand}, {Benetti},
  {Elias-Rosa}, {Fraser}, {Gal-Yam}, {Hamuy}, {Inserra}, {Knapic}, {LaCluyze},
  {Molinaro}, {Ochner}, {Pastorello}, {Pignata}, {Reichart}, {Ries},
  {Riffeser}, {Schmidt}, {Schmidt}, {Smareglia}, {Smartt}, {Smith},
  {Sollerman}, {Sullivan}, {Tomasella}, {Turatto}, {Valenti}, {Yaron}, \&
  {Young}}]{Barbarino2015}
{Barbarino}, C., {Dall'Ora}, M., {Botticella}, M.~T., {et~al.} 2015, \mnras,
  448, 2312

\bibitem[{{Beck}(2015)}]{Beck2015}
{Beck}, R. 2015, \aapr, 24, 4

\bibitem[{{Bessell}(1990)}]{Bessell1990}
{Bessell}, M.~S. 1990, \pasp, 102, 1181

\bibitem[{{Blondin} {et~al.}(2006){Blondin}, {Modjaz}, {Kirshner}, {Challis},
  \& {Berlind}}]{Blondin2006}
{Blondin}, S., {Modjaz}, M., {Kirshner}, R., {Challis}, P., \& {Berlind}, P.
  2006, Central Bureau Electronic Telegrams, 757, 1

\bibitem[{{Bose} {et~al.}(2013){Bose}, {Kumar}, {Sutaria}, {Kumar}, {Roy},
  {Bhatt}, {Pandey}, {Chandola}, {Sagar}, {Misra}, \& {Chakraborti}}]{Bose2013}
{Bose}, S., {Kumar}, B., {Sutaria}, F., {et~al.} 2013, \mnras, 433, 1871

\bibitem[{{Bose} {et~al.}(2015){Bose}, {Sutaria}, {Kumar}, {Duggal}, {Misra},
  {Brown}, {Singh}, {Dwarkadas}, {York}, {Chakraborti}, {Chandola},
  {Dahlstrom}, {Ray}, \& {Safonova}}]{Bose2015}
{Bose}, S., {Sutaria}, F., {Kumar}, B., {et~al.} 2015, \apj, 806, 160

\bibitem[{{Brown} {et~al.}(2014){Brown}, {Breeveld}, {Holland}, {Kuin}, \&
  {Pritchard}}]{Brown2014}
{Brown}, P.~J., {Breeveld}, A.~A., {Holland}, S., {Kuin}, P., \& {Pritchard},
  T. 2014, \apss, 354, 89

\bibitem[{{Chassagne}(2001)}]{Chassagne2001}
{Chassagne}, R. 2001, \iaucirc, 7670, 3

\bibitem[{{Chornock} {et~al.}(2006){Chornock}, {Filippenko}, {Branch}, {Foley},
  {Jha}, \& {Li}}]{Chornock2006}
{Chornock}, R., {Filippenko}, A.~V., {Branch}, D., {et~al.} 2006, \pasp, 118,
  722

\bibitem[{{Chornock} {et~al.}(2010){Chornock}, {Filippenko}, {Li}, \&
  {Silverman}}]{Chornock2010}
{Chornock}, R., {Filippenko}, A.~V., {Li}, W., \& {Silverman}, J.~M. 2010,
  \apj, 713, 1363

\bibitem[{{Chu} {et~al.}(2022){Chu}, {Cikota}, {Baade}, {Patat}, {Filippenko},
  {Wheeler}, {Maund}, {Bulla}, {Yang}, {H{\"o}flich}, \& {Wang}}]{Chu2022}
{Chu}, M.~R., {Cikota}, A., {Baade}, D., {et~al.} 2022, \mnras, 509, 6028

\bibitem[{{Chugai} {et~al.}(2005){Chugai}, {Fabrika}, {Sholukhova},
  {Goranskij}, {Abolmasov}, \& {Vlasyuk}}]{Chugai2005}
{Chugai}, N.~N., {Fabrika}, S.~N., {Sholukhova}, O.~N., {et~al.} 2005,
  Astronomy Letters, 31, 792

\bibitem[{{Cikota} {et~al.}(2018){Cikota}, {Hoang}, {Taubenberger}, {Patat},
  {Mazzei}, {Cox}, {Zelaya}, {Cikota}, {Tomasella}, {Benetti}, \&
  {Rodeghiero}}]{Cikota2018}
{Cikota}, A., {Hoang}, T., {Taubenberger}, S., {et~al.} 2018, \aap, 615, A42

\bibitem[{{Dall'Ora} {et~al.}(2014){Dall'Ora}, {Botticella}, {Pumo},
  {Zampieri}, {Tomasella}, {Pignata}, {Bayless}, {Pritchard}, {Taubenberger},
  {Kotak}, {Inserra}, {Della Valle}, {Cappellaro}, {Benetti}, {Benitez},
  {Bufano}, {Elias-Rosa}, {Fraser}, {Haislip}, {Harutyunyan}, {Howell},
  {Hsiao}, {Iijima}, {Kankare}, {Kuin}, {Maund}, {Morales-Garoffolo},
  {Morrell}, {Munari}, {Ochner}, {Pastorello}, {Patat}, {Phillips}, {Reichart},
  {Roming}, {Siviero}, {Smartt}, {Sollerman}, {Taddia}, {Valenti}, \&
  {Wright}}]{DallOra2014}
{Dall'Ora}, M., {Botticella}, M.~T., {Pumo}, M.~L., {et~al.} 2014, \apj, 787,
  139

\bibitem[{{Davis} \& {Greenstein}(1951)}]{Davis1951}
{Davis}, Leverett, J. \& {Greenstein}, J.~L. 1951, \apj, 114, 206

\bibitem[{{de Jaeger} {et~al.}(2019){de Jaeger}, {Zheng}, {Stahl},
  {Filippenko}, {Brink}, {Bigley}, {Blanchard}, {Blanchard}, {Bradley},
  {Cargill}, {Casper}, {Cenko}, {Channa}, {Choi}, {Clubb}, {Cobb}, {Cohen}, {de
  Kouchkovsky}, {Ellison}, {Falcon}, {Fox}, {Fuller}, {Ganeshalingam}, {Gould},
  {Graham}, {Halevi}, {Hayakawa}, {Hestenes}, {Hyland}, {Jeffers}, {Joubert},
  {Kandrashoff}, {Kelly}, {Kim}, {Kim}, {Kumar}, {Leonard}, {Li}, {Lowe}, {Lu},
  {Mason}, {McAllister}, {Mauerhan}, {Modjaz}, {Molloy}, {Perley}, {Pina},
  {Poznanski}, {Ross}, {Shivvers}, {Silverman}, {Soler}, {Stegman}, {Taylor},
  {Tang}, {Wilkins}, {Wang}, {Wang}, {Yuk}, {Yunus}, \& {Zhang}}]{deJaeger2019}
{de Jaeger}, T., {Zheng}, W., {Stahl}, B.~E., {et~al.} 2019, \mnras, 490, 2799

\bibitem[{{Deng} {et~al.}(1999){Deng}, {Cao}, {Xu}, {Lu}, {Li}, {Qiu}, \&
  {Hu}}]{Deng1999}
{Deng}, J.~S., {Cao}, H.~L., {Xu}, D.~W., {et~al.} 1999, \iaucirc, 7296, 3

\bibitem[{{Dessart} {et~al.}(2021){Dessart}, {Leonard}, {John Hillier}, \&
  {Pignata}}]{Dessart2021}
{Dessart}, L., {Leonard}, D.~C., {John Hillier}, D., \& {Pignata}, G. 2021,
  \aap, 651, A19

\bibitem[{{Dhungana} {et~al.}(2016){Dhungana}, {Kehoe}, {Vinko}, {Silverman},
  {Wheeler}, {Zheng}, {Marion}, {Fox}, {Akerlof}, {Biro}, {Borkovits}, {Cenko},
  {Clubb}, {Filippenko}, {Ferrante}, {Gibson}, {Graham}, {Hegedus}, {Kelly},
  {Kelemen}, {Lee}, {Marschalko}, {Moln{\'a}r}, {Nagy}, {Ordasi}, {Pal},
  {Sarneczky}, {Shivvers}, {Szakats}, {Szalai}, {Szegedi-Elek}, {Sz{\'e}kely},
  {Szing}, {Tak{\'a}ts}, \& {Vida}}]{Dhungana2016}
{Dhungana}, G., {Kehoe}, R., {Vinko}, J., {et~al.} 2016, \apj, 822, 6

\bibitem[{{Doi} {et~al.}(2007){Doi}, {Nakano}, {Itagaki}, {Naito}, \&
  {Iizuka}}]{Doi2007}
{Doi}, T., {Nakano}, S., {Itagaki}, K., {Naito}, H., \& {Iizuka}, R. 2007,
  Central Bureau Electronic Telegrams, 848, 1

\bibitem[{{Evans} {et~al.}(2001){Evans}, {Bock}, {Marples}, {Garradd}, {Salvo},
  {Schmidt}, {Ashley}, {Phillips}, {Stubbs}, {Francis}, \& {Hale}}]{Evans2001}
{Evans}, R., {Bock}, G., {Marples}, P., {et~al.} 2001, \iaucirc, 7690, 1

\bibitem[{{Fagotti} {et~al.}(2012){Fagotti}, {Dimai}, {Quadri}, {Strabla},
  {Girelli}, {Quadri}, {Fiorentino}, {Skvarc}, \& {Masi}}]{Fagotti2012}
{Fagotti}, P., {Dimai}, A., {Quadri}, U., {et~al.} 2012, Central Bureau
  Electronic Telegrams, 3054, 1

\bibitem[{{Folatelli} {et~al.}(2007){Folatelli}, {Gonzalez}, \&
  {Morrell}}]{Folatelli2007}
{Folatelli}, G., {Gonzalez}, S., \& {Morrell}, N. 2007, Central Bureau
  Electronic Telegrams, 850, 1

\bibitem[{{Fraser} {et~al.}(2012){Fraser}, {Kotak}, {Wright}, {Smith},
  {Smartt}, {Magill}, {McCrum}, \& {Jensen-Hansen}}]{Fraser2012}
{Fraser}, M., {Kotak}, R., {Wright}, D., {et~al.} 2012, Central Bureau
  Electronic Telegrams, 3173, 1

\bibitem[{{Galbany} {et~al.}(2016){Galbany}, {Hamuy}, {Phillips}, {Suntzeff},
  {Maza}, {de Jaeger}, {Moraga}, {Gonz{\'a}lez-Gait{\'a}n}, {Krisciunas},
  {Morrell}, {Thomas-Osip}, {Krzeminski}, {Gonz{\'a}lez}, {Antezana},
  {Wishnjewski}, {McCarthy}, {Anderson}, {Guti{\'e}rrez}, {Stritzinger},
  {Folatelli}, {Anguita}, {Galaz}, {Green}, {Impey}, {Kim}, {Kirhakos},
  {Malkan}, {Mulchaey}, {Phillips}, {Pizzella}, {Prosser}, {Schmidt},
  {Schommer}, {Sherry}, {Strolger}, {Wells}, \& {Williger}}]{Galbany2016}
{Galbany}, L., {Hamuy}, M., {Phillips}, M.~M., {et~al.} 2016, \aj, 151, 33

\bibitem[{{Green}(2010)}]{Green2010}
{Green}, D.~W.~E. 2010, Central Bureau Electronic Telegrams, 2293, 2

\bibitem[{{Guillochon} {et~al.}(2017){Guillochon}, {Parrent}, {Kelley}, \&
  {Margutti}}]{Guillochon2017}
{Guillochon}, J., {Parrent}, J., {Kelley}, L.~Z., \& {Margutti}, R. 2017, \apj,
  835, 64

\bibitem[{{Guti{\'e}rrez} {et~al.}(2017){Guti{\'e}rrez}, {Anderson}, {Hamuy},
  {Morrell}, {Gonz{\'a}lez-Gaitan}, {Stritzinger}, {Phillips}, {Galbany},
  {Folatelli}, {Dessart}, {Contreras}, {Della Valle}, {Freedman}, {Hsiao},
  {Krisciunas}, {Madore}, {Maza}, {Suntzeff}, {Prieto}, {Gonz{\'a}lez},
  {Cappellaro}, {Navarrete}, {Pizzella}, {Ruiz}, {Smith}, \&
  {Turatto}}]{Gutierrez2017}
{Guti{\'e}rrez}, C.~P., {Anderson}, J.~P., {Hamuy}, M., {et~al.} 2017, \apj,
  850, 89

\bibitem[{{Hosseinzadeh} {et~al.}(2017){Hosseinzadeh}, {Valenti}, {Arcavi},
  {Howell}, {McCully}, {Sand}, \& {Tartaglia}}]{Hosseinzadeh2017}
{Hosseinzadeh}, G., {Valenti}, S., {Arcavi}, I., {et~al.} 2017, The
  Astronomer's Telegram, 10059, 1

\bibitem[{{Howell} {et~al.}(2001){Howell}, {H{\"o}flich}, {Wang}, \&
  {Wheeler}}]{Howell2001}
{Howell}, D.~A., {H{\"o}flich}, P., {Wang}, L., \& {Wheeler}, J.~C. 2001, \apj,
  556, 302

\bibitem[{{Huang} {et~al.}(2015){Huang}, {Wang}, {Zhang}, {Brown}, {Zampieri},
  {Pumo}, {Zhang}, {Chen}, {Mo}, \& {Zhao}}]{Huang2015}
{Huang}, F., {Wang}, X., {Zhang}, J., {et~al.} 2015, \apj, 807, 59

\bibitem[{{Inserra} {et~al.}(2016){Inserra}, {Bulla}, {Sim}, \&
  {Smartt}}]{Inserra2016}
{Inserra}, C., {Bulla}, M., {Sim}, S.~A., \& {Smartt}, S.~J. 2016, \apj, 831,
  79

\bibitem[{{Itoh} {et~al.}(2012){Itoh}, {Ui}, \& {Yamanaka}}]{Itoh2012}
{Itoh}, R., {Ui}, T., \& {Yamanaka}, M. 2012, Central Bureau Electronic
  Telegrams, 3054, 2

\bibitem[{{Kim} {et~al.}(2013){Kim}, {Zheng}, {Li}, {Filippenko}, {Cenko},
  {Richmond}, {Amorim}, {Balam}, {Graham}, \& {Hsiao}}]{Kim2013}
{Kim}, M., {Zheng}, W., {Li}, W., {et~al.} 2013, Central Bureau Electronic
  Telegrams, 3606, 1

\bibitem[{{Koribalski} {et~al.}(2004){Koribalski}, {Staveley-Smith}, {Kilborn},
  {Ryder}, {Kraan-Korteweg}, {Ryan-Weber}, {Ekers}, {Jerjen}, {Henning},
  {Putman}, {Zwaan}, {de Blok}, {Calabretta}, {Disney}, {Minchin}, {Bhathal},
  {Boyce}, {Drinkwater}, {Freeman}, {Gibson}, {Green}, {Haynes}, {Juraszek},
  {Kesteven}, {Knezek}, {Mader}, {Marquarding}, {Meyer}, {Mould}, {Oosterloo},
  {O'Brien}, {Price}, {Sadler}, {Schr{\"o}der}, {Stewart}, {Stootman}, {Waugh},
  {Warren}, {Webster}, \& {Wright}}]{Koribalski2004}
{Koribalski}, B.~S., {Staveley-Smith}, L., {Kilborn}, V.~A., {et~al.} 2004,
  \aj, 128, 16

\bibitem[{{Leonard} {et~al.}(2012){Leonard}, {Dessart}, {Hillier}, \&
  {Pignata}}]{Leonard2012}
{Leonard}, D.~C., {Dessart}, L., {Hillier}, D.~J., \& {Pignata}, G. 2012, in
  American Institute of Physics Conference Series, Vol. 1429, Stellar
  Polarimetry: from Birth to Death, ed. J.~L. {Hoffman}, J.~{Bjorkman}, \&
  B.~{Whitney}, 204--207

\bibitem[{{Leonard} {et~al.}(2021){Leonard}, {Dessart}, {Hillier}, {Pignata},
  {Williams}, {Hoffman}, {Milne}, {Smith}, {Smith}, \&
  {Khandrika}}]{Leonard2021}
{Leonard}, D.~C., {Dessart}, L., {Hillier}, D.~J., {et~al.} 2021, \apjl, 921,
  L35

\bibitem[{{Leonard} {et~al.}(2001){Leonard}, {Filippenko}, {Ardila}, \&
  {Brotherton}}]{Leonard2001}
{Leonard}, D.~C., {Filippenko}, A.~V., {Ardila}, D.~R., \& {Brotherton}, M.~S.
  2001, \apj, 553, 861

\bibitem[{{Leonard} {et~al.}(2000){Leonard}, {Filippenko}, {Barth}, \&
  {Matheson}}]{Leonard2000}
{Leonard}, D.~C., {Filippenko}, A.~V., {Barth}, A.~J., \& {Matheson}, T. 2000,
  \apj, 536, 239

\bibitem[{{Leonard} {et~al.}(2002{\natexlab{a}}){Leonard}, {Filippenko},
  {Chornock}, \& {Foley}}]{Leonard2002b}
{Leonard}, D.~C., {Filippenko}, A.~V., {Chornock}, R., \& {Foley}, R.~J.
  2002{\natexlab{a}}, \pasp, 114, 1333

\bibitem[{{Leonard} {et~al.}(2006){Leonard}, {Filippenko}, {Ganeshalingam},
  {Serduke}, {Li}, {Swift}, {Gal-Yam}, {Foley}, {Fox}, {Park}, {Hoffman}, \&
  {Wong}}]{Leonard2006}
{Leonard}, D.~C., {Filippenko}, A.~V., {Ganeshalingam}, M., {et~al.} 2006,
  \nat, 440, 505

\bibitem[{{Leonard} {et~al.}(2002{\natexlab{b}}){Leonard}, {Filippenko},
  {Gates}, {Li}, {Eastman}, {Barth}, {Bus}, {Chornock}, {Coil}, {Frink},
  {Grady}, {Harris}, {Malkan}, {Matheson}, {Quirrenbach}, \&
  {Treffers}}]{Leonard2002a}
{Leonard}, D.~C., {Filippenko}, A.~V., {Gates}, E.~L., {et~al.}
  2002{\natexlab{b}}, \pasp, 114, 35

\bibitem[{{Leonard} {et~al.}(2003){Leonard}, {Kanbur}, {Ngeow}, \&
  {Tanvir}}]{Leonard2003}
{Leonard}, D.~C., {Kanbur}, S.~M., {Ngeow}, C.~C., \& {Tanvir}, N.~R. 2003,
  \apj, 594, 247

\bibitem[{{Leonard} {et~al.}(2005){Leonard}, {Li}, {Filippenko}, {Foley}, \&
  {Chornock}}]{Leonard2005}
{Leonard}, D.~C., {Li}, W., {Filippenko}, A.~V., {Foley}, R.~J., \& {Chornock},
  R. 2005, \apj, 632, 450

\bibitem[{{Li} {et~al.}(2007){Li}, {Wang}, {Van Dyk}, {Cuillandre}, {Foley}, \&
  {Filippenko}}]{Li2007}
{Li}, W., {Wang}, X., {Van Dyk}, S.~D., {et~al.} 2007, \apj, 661, 1013

\bibitem[{{Li}(1999)}]{Li1999}
{Li}, W.~D. 1999, \iaucirc, 7294, 1

\bibitem[{{Lu} {et~al.}(1993){Lu}, {Hoffman}, {Groff}, {Roos}, \&
  {Lamphier}}]{Lu1993}
{Lu}, N.~Y., {Hoffman}, G.~L., {Groff}, T., {Roos}, T., \& {Lamphier}, C. 1993,
  \apjs, 88, 383

\bibitem[{{Mauerhan} {et~al.}(2015){Mauerhan}, {Williams}, {Leonard}, {Smith},
  {Filippenko}, {Smith}, {Hoffman}, {Huk}, {Clubb}, {Silverman}, {Cenko},
  {Milne}, {Gal-Yam}, \& {Ben-Ami}}]{Mauerhan2015}
{Mauerhan}, J.~C., {Williams}, G.~G., {Leonard}, D.~C., {et~al.} 2015, \mnras,
  453, 4467

\bibitem[{{Maza} {et~al.}(2012){Maza}, {Hamuy}, {Antezana}, {Gonzalez},
  {Cartier}, {Forster}, {Silva}, {Carrasco}, {Sanchez}, {Hervias}, {Ramirez},
  {Pignata}, {Cifuentes}, {Apostolovski}, {Vidal}, {Aros}, {Conuel},
  {Folatelli}, {Reichart}, {Ivarsen}, {Haislip}, {Crain}, {Foster},
  {Nysewander}, {LaCluyze}, {Milisavljevic}, {Fesen}, {Soderberg}, {Margutti},
  {Pickering}, {Kotze}, \& {Crawford}}]{Maza2012}
{Maza}, J., {Hamuy}, M., {Antezana}, R., {et~al.} 2012, Central Bureau
  Electronic Telegrams, 3163, 1

\bibitem[{{Monard}(2008)}]{Monard2008}
{Monard}, L.~A.~G. 2008, Central Bureau Electronic Telegrams, 1315, 1

\bibitem[{{Monard}(2010)}]{Monard2010}
{Monard}, L.~A.~G. 2010, Central Bureau Electronic Telegrams, 2271, 1

\bibitem[{{Monard} {et~al.}(2012){Monard}, {Childress}, {Scalzo}, {Yuan}, \&
  {Schmidt}}]{Monard2012}
{Monard}, L.~A.~G., {Childress}, M., {Scalzo}, R., {Yuan}, F., \& {Schmidt}, B.
  2012, Central Bureau Electronic Telegrams, 3201, 1

\bibitem[{{Morrell}(2010)}]{Morrell2010}
{Morrell}, N. 2010, Central Bureau Electronic Telegrams, 2293, 1

\bibitem[{{Morrell} \& {Stritzinger}(2008)}]{Morrell2008}
{Morrell}, N. \& {Stritzinger}, M. 2008, Central Bureau Electronic Telegrams,
  1335, 1

\bibitem[{{Munari} {et~al.}(2013){Munari}, {Henden}, {Belligoli}, {Castellani},
  {Cherini}, {Righetti}, \& {Vagnozzi}}]{Munari2013}
{Munari}, U., {Henden}, A., {Belligoli}, R., {et~al.} 2013, \na, 20, 30

\bibitem[{{Nagao} {et~al.}(2019){Nagao}, {Cikota}, {Patat}, {Taubenberger},
  {Bulla}, {Faran}, {Sand}, {Valenti}, {Andrews}, \& {Reichart}}]{Nagao2019}
{Nagao}, T., {Cikota}, A., {Patat}, F., {et~al.} 2019, \mnras, 489, L69

\bibitem[{{Nagao} {et~al.}(2022){Nagao}, {Patat}, {Maeda}, {Baade}, {Mattila},
  {Taubenberger}, {Kotak}, {Cikota}, {Kuncarayakti}, {Bulla}, \&
  {Maund}}]{Nagao2022}
{Nagao}, T., {Patat}, F., {Maeda}, K., {et~al.} 2022, \apjl, 941, L4

\bibitem[{{Nagao} {et~al.}(2021){Nagao}, {Patat}, {Taubenberger}, {Baade},
  {Faran}, {Cikota}, {Sand}, {Bulla}, {Kuncarayakti}, {Maund}, {Tartaglia},
  {Valenti}, \& {Reichart}}]{Nagao2021}
{Nagao}, T., {Patat}, F., {Taubenberger}, S., {et~al.} 2021, \mnras, 505, 3664

\bibitem[{{Nakano} {et~al.}(2004){Nakano}, {Itagaki}, {Bouma}, {Lehky}, \&
  {Hornoch}}]{Nakano2004}
{Nakano}, S., {Itagaki}, K., {Bouma}, R.~J., {Lehky}, M., \& {Hornoch}, K.
  2004, \iaucirc, 8377, 1

\bibitem[{{Nakano} {et~al.}(2006){Nakano}, {Itagaki}, \& {Kadota}}]{Nakano2006}
{Nakano}, S., {Itagaki}, K., \& {Kadota}, K. 2006, Central Bureau Electronic
  Telegrams, 756, 1

\bibitem[{{Ntormousi}(2018)}]{Ntormousi2018}
{Ntormousi}, E. 2018, \aap, 619, L5

\bibitem[{{Patat} {et~al.}(2004){Patat}, {Benetti}, {Pastorello}, {Filippenko},
  \& {Aceituno}}]{Patat2004}
{Patat}, F., {Benetti}, S., {Pastorello}, A., {Filippenko}, A.~V., \&
  {Aceituno}, J. 2004, \iaucirc, 8378, 1

\bibitem[{{Patat} {et~al.}(2001){Patat}, {Contreras}, {Prieto}, {Altavilla},
  {Benetti}, {Cappellaro}, {Pastorello}, \& {Turatto}}]{Patat2001}
{Patat}, F., {Contreras}, C., {Prieto}, J., {et~al.} 2001, \iaucirc, 7680, 1

\bibitem[{{Patat} {et~al.}(2012){Patat}, {H{\"o}flich}, {Baade}, {Maund},
  {Wang}, \& {Wheeler}}]{Patat2012}
{Patat}, F., {H{\"o}flich}, P., {Baade}, D., {et~al.} 2012, \aap, 545, A7

\bibitem[{{Patat} \& {Romaniello}(2006)}]{Patat2006}
{Patat}, F. \& {Romaniello}, M. 2006, \pasp, 118, 146

\bibitem[{{Patat} {et~al.}(2015){Patat}, {Taubenberger}, {Cox}, {Baade},
  {Clocchiatti}, {H{\"o}flich}, {Maund}, {Reilly}, {Spyromilio}, {Wang},
  {Wheeler}, \& {Zelaya}}]{Patat2015}
{Patat}, F., {Taubenberger}, S., {Cox}, N.~L.~J., {et~al.} 2015, \aap, 577, A53

\bibitem[{{Pignata} {et~al.}(2010){Pignata}, {Cifuentes}, {Maza}, {Hamuy},
  {Antezana}, {Gonzalez}, {Gonzalez}, {Silva}, {Folatelli}, {Cartier},
  {Forster}, {Marchi}, {Conuel}, {Reichart}, {Ivarsen}, {Haislip}, {Crain},
  {Foster}, {Nysewander}, {Lacluyze}, \& {Morrell}}]{Pignata2010}
{Pignata}, G., {Cifuentes}, M., {Maza}, J., {et~al.} 2010, Central Bureau
  Electronic Telegrams, 2456, 1

\bibitem[{{Poznanski} {et~al.}(2012){Poznanski}, {Nugent}, {Ofek}, {Gal-Yam},
  \& {Kasliwal}}]{Poznanski2012}
{Poznanski}, D., {Nugent}, P.~E., {Ofek}, E.~O., {Gal-Yam}, A., \& {Kasliwal},
  M.~M. 2012, The Astronomer's Telegram, 3996, 1

\bibitem[{{Pursimo} {et~al.}(2017){Pursimo}, {Elias-Rosa}, {Dennefeld},
  {Korhonen}, {Cabezas}, {NEON School PhD Students}, {Fernandes}, {McWhirter},
  \& {Zervas}}]{Pursimo2017}
{Pursimo}, T., {Elias-Rosa}, N., {Dennefeld}, M., {et~al.} 2017, The
  Astronomer's Telegram, 10717, 1

\bibitem[{{Reilly} {et~al.}(2017){Reilly}, {Maund}, {Baade}, {Wheeler},
  {H{\"o}flich}, {Spyromilio}, {Patat}, \& {Wang}}]{Reilly2017}
{Reilly}, E., {Maund}, J.~R., {Baade}, D., {et~al.} 2017, \mnras, 470, 1491

\bibitem[{{Schlafly} \& {Finkbeiner}(2011)}]{Schlafly2011}
{Schlafly}, E.~F. \& {Finkbeiner}, D.~P. 2011, \apj, 737, 103

\bibitem[{{Serkowski} {et~al.}(1975){Serkowski}, {Mathewson}, \&
  {Ford}}]{Serkowski1975}
{Serkowski}, K., {Mathewson}, D.~S., \& {Ford}, V.~L. 1975, \apj, 196, 261

\bibitem[{{Shappee} {et~al.}(2013){Shappee}, {Kochanek}, {Stanek}, {Basu},
  {Holoien}, {Jencson}, {Beacom}, {Prieto}, {Szczygiel}, {Pojmanski},
  {Dubberley}, {Elphick}, {Foale}, {Hawkins}, {Mullens}, {Rosing}, {Ross},
  {Walker}, \& {Brimacombe}}]{Shappee2013}
{Shappee}, B.~J., {Kochanek}, C.~S., {Stanek}, K.~Z., {et~al.} 2013, The
  Astronomer's Telegram, 5237, 1

\bibitem[{{Silverman} {et~al.}(2012){Silverman}, {Foley}, {Filippenko},
  {Ganeshalingam}, {Barth}, {Chornock}, {Griffith}, {Kong}, {Lee}, {Leonard},
  {Matheson}, {Miller}, {Steele}, {Barris}, {Bloom}, {Cobb}, {Coil},
  {Desroches}, {Gates}, {Ho}, {Jha}, {Kandrashoff}, {Li}, {Mandel}, {Modjaz},
  {Moore}, {Mostardi}, {Papenkova}, {Park}, {Perley}, {Poznanski}, {Reuter},
  {Scala}, {Serduke}, {Shields}, {Swift}, {Tonry}, {Van Dyk}, {Wang}, \&
  {Wong}}]{Silverman2012}
{Silverman}, J.~M., {Foley}, R.~J., {Filippenko}, A.~V., {et~al.} 2012, \mnras,
  425, 1789

\bibitem[{{Smartt} {et~al.}(2001){Smartt}, {Kilkenny}, \&
  {Meikle}}]{Smartt2001}
{Smartt}, S.~J., {Kilkenny}, D., \& {Meikle}, P. 2001, \iaucirc, 7704, 1

\bibitem[{{Smartt} {et~al.}(2015){Smartt}, {Valenti}, {Fraser}, {Inserra},
  {Young}, {Sullivan}, {Pastorello}, {Benetti}, {Gal-Yam}, {Knapic},
  {Molinaro}, {Smareglia}, {Smith}, {Taubenberger}, {Yaron}, {Anderson},
  {Ashall}, {Balland}, {Baltay}, {Barbarino}, {Bauer}, {Baumont}, {Bersier},
  {Blagorodnova}, {Bongard}, {Botticella}, {Bufano}, {Bulla}, {Cappellaro},
  {Campbell}, {Cellier-Holzem}, {Chen}, {Childress}, {Clocchiatti},
  {Contreras}, {Dall'Ora}, {Danziger}, {de Jaeger}, {De Cia}, {Della Valle},
  {Dennefeld}, {Elias-Rosa}, {Elman}, {Feindt}, {Fleury}, {Gall},
  {Gonzalez-Gaitan}, {Galbany}, {Morales Garoffolo}, {Greggio}, {Guillou},
  {Hachinger}, {Hadjiyska}, {Hage}, {Hillebrandt}, {Hodgkin}, {Hsiao}, {James},
  {Jerkstrand}, {Kangas}, {Kankare}, {Kotak}, {Kromer}, {Kuncarayakti},
  {Leloudas}, {Lundqvist}, {Lyman}, {Hook}, {Maguire}, {Manulis}, {Margheim},
  {Mattila}, {Maund}, {Mazzali}, {McCrum}, {McKinnon}, {Moreno-Raya},
  {Nicholl}, {Nugent}, {Pain}, {Pignata}, {Phillips}, {Polshaw}, {Pumo},
  {Rabinowitz}, {Reilly}, {Romero-Ca{\~n}izales}, {Scalzo}, {Schmidt},
  {Schulze}, {Sim}, {Sollerman}, {Taddia}, {Tartaglia}, {Terreran},
  {Tomasella}, {Turatto}, {Walker}, {Walton}, {Wyrzykowski}, {Yuan}, \&
  {Zampieri}}]{Smartt2015}
{Smartt}, S.~J., {Valenti}, S., {Fraser}, M., {et~al.} 2015, \aap, 579, A40

\bibitem[{{Sorce} {et~al.}(2014){Sorce}, {Tully}, {Courtois}, {Jarrett},
  {Neill}, \& {Shaya}}]{Sorce2014}
{Sorce}, J.~G., {Tully}, R.~B., {Courtois}, H.~M., {et~al.} 2014, \mnras, 444,
  527

\bibitem[{{Spiro} {et~al.}(2014){Spiro}, {Pastorello}, {Pumo}, {Zampieri},
  {Turatto}, {Smartt}, {Benetti}, {Cappellaro}, {Valenti}, {Agnoletto},
  {Altavilla}, {Aoki}, {Brocato}, {Corsini}, {Di Cianno}, {Elias-Rosa},
  {Hamuy}, {Enya}, {Fiaschi}, {Folatelli}, {Desidera}, {Harutyunyan}, {Howell},
  {Kawka}, {Kobayashi}, {Leibundgut}, {Minezaki}, {Navasardyan}, {Nomoto},
  {Mattila}, {Pietrinferni}, {Pignata}, {Raimondo}, {Salvo}, {Schmidt},
  {Sollerman}, {Spyromilio}, {Taubenberger}, {Valentini}, {Vennes}, \&
  {Yoshii}}]{Spiro2014}
{Spiro}, S., {Pastorello}, A., {Pumo}, M.~L., {et~al.} 2014, \mnras, 439, 2873

\bibitem[{{Springob} {et~al.}(2009){Springob}, {Masters}, {Haynes},
  {Giovanelli}, \& {Marinoni}}]{Springob2009}
{Springob}, C.~M., {Masters}, K.~L., {Haynes}, M.~P., {Giovanelli}, R., \&
  {Marinoni}, C. 2009, \apjs, 182, 474

\bibitem[{{Stevance} {et~al.}(2019){Stevance}, {Maund}, {Baade}, {Bruten},
  {Cikota}, {H{\"o}flich}, {Wang}, {Wheeler}, {Clocchiatti}, {Spyromilio},
  {Patat}, {Yang}, \& {Crowther}}]{Stevance2019}
{Stevance}, H.~F., {Maund}, J.~R., {Baade}, D., {et~al.} 2019, \mnras, 485, 102

\bibitem[{{Stevance} {et~al.}(2017){Stevance}, {Maund}, {Baade}, {H{\"o}flich},
  {Howerton}, {Patat}, {Rose}, {Spyromilio}, {Wheeler}, \&
  {Wang}}]{Stevance2017}
{Stevance}, H.~F., {Maund}, J.~R., {Baade}, D., {et~al.} 2017, \mnras, 469,
  1897

\bibitem[{{Takaki} {et~al.}(2012){Takaki}, {Moritani}, {Itoh}, {Ueno}, {Urano},
  {Kawabata}, \& {Yamanaka}}]{Takaki2012}
{Takaki}, K., {Moritani}, Y., {Itoh}, R., {et~al.} 2012, Central Bureau
  Electronic Telegrams, 3203, 1

\bibitem[{{Tak{\'a}ts} \& {Vink{\'o}}(2012)}]{Takats2012}
{Tak{\'a}ts}, K. \& {Vink{\'o}}, J. 2012, \mnras, 419, 2783

\bibitem[{{Tartaglia} {et~al.}(2017){Tartaglia}, {Sand}, {Valenti}, {Reichart},
  {Haislip}, \& {Kouprianov}}]{Tartaglia2017}
{Tartaglia}, L., {Sand}, D., {Valenti}, S., {et~al.} 2017, The Astronomer's
  Telegram, 10058, 1

\bibitem[{{Tartaglia} {et~al.}(2021){Tartaglia}, {Sand}, {Groh}, {Valenti},
  {Wyatt}, {Bostroem}, {Brown}, {Yang}, {Burke}, {Chen}, {Davis},
  {F{\"o}rster}, {Galbany}, {Haislip}, {Hiramatsu}, {Hosseinzadeh}, {Howell},
  {Hsiao}, {Jha}, {Kouprianov}, {Kuncarayakti}, {Lyman}, {McCully}, {Phillips},
  {Rau}, {Reichart}, {Shahbandeh}, \& {Strader}}]{Tartaglia2021}
{Tartaglia}, L., {Sand}, D.~J., {Groh}, J.~H., {et~al.} 2021, \apj, 907, 52

\bibitem[{{Tartaglia} {et~al.}(2018){Tartaglia}, {Sand}, {Valenti}, {Wyatt},
  {Anderson}, {Arcavi}, {Ashall}, {Botticella}, {Cartier}, {Chen}, {Cikota},
  {Coulter}, {Della Valle}, {Foley}, {Gal-Yam}, {Galbany}, {Gall}, {Haislip},
  {Harmanen}, {Hosseinzadeh}, {Howell}, {Hsiao}, {Inserra}, {Jha}, {Kankare},
  {Kilpatrick}, {Kouprianov}, {Kuncarayakti}, {Maccarone}, {Maguire},
  {Mattila}, {Mazzali}, {McCully}, {Melandri}, {Morrell}, {Phillips},
  {Pignata}, {Piro}, {Prentice}, {Reichart}, {Rojas-Bravo}, {Smartt}, {Smith},
  {Sollerman}, {Stritzinger}, {Sullivan}, {Taddia}, \& {Young}}]{Tartaglia2018}
{Tartaglia}, L., {Sand}, D.~J., {Valenti}, S., {et~al.} 2018, \apj, 853, 62

\bibitem[{{Tody}(1986)}]{Tody1986}
{Tody}, D. 1986, in Society of Photo-Optical Instrumentation Engineers (SPIE)
  Conference Series, Vol. 627, Instrumentation in astronomy VI, ed. D.~L.
  {Crawford}, 733

\bibitem[{{Tody}(1993)}]{Tody1993}
{Tody}, D. 1993, in Astronomical Society of the Pacific Conference Series,
  Vol.~52, Astronomical Data Analysis Software and Systems II, ed. R.~J.
  {Hanisch}, R.~J.~V. {Brissenden}, \& J.~{Barnes}, 173

\bibitem[{{Trammell} {et~al.}(1993){Trammell}, {Hines}, \&
  {Wheeler}}]{Trammell1993}
{Trammell}, S.~R., {Hines}, D.~C., \& {Wheeler}, J.~C. 1993, \apjl, 414, L21

\bibitem[{{Tran} {et~al.}(1997){Tran}, {Filippenko}, {Schmidt}, {Bjorkman},
  {Jannuzi}, \& {Smith}}]{Tran1997}
{Tran}, H.~D., {Filippenko}, A.~V., {Schmidt}, G.~D., {et~al.} 1997, \pasp,
  109, 489

\bibitem[{{Tsvetkov} {et~al.}(2008){Tsvetkov}, {Goranskij}, \&
  {Pavlyuk}}]{Tsvetkov2008}
{Tsvetkov}, D.~Y., {Goranskij}, V., \& {Pavlyuk}, N. 2008, Peremennye Zvezdy,
  28, 8

\bibitem[{{Tully} {et~al.}(2013){Tully}, {Courtois}, {Dolphin}, {Fisher},
  {H{\'e}raudeau}, {Jacobs}, {Karachentsev}, {Makarov}, {Makarova},
  {Mitronova}, {Rizzi}, {Shaya}, {Sorce}, \& {Wu}}]{Tully2013}
{Tully}, R.~B., {Courtois}, H.~M., {Dolphin}, A.~E., {et~al.} 2013, \aj, 146,
  86

\bibitem[{{Tully} {et~al.}(2016){Tully}, {Courtois}, \& {Sorce}}]{Tully2016}
{Tully}, R.~B., {Courtois}, H.~M., \& {Sorce}, J.~G. 2016, \aj, 152, 50

\bibitem[{{Valenti} {et~al.}(2017){Valenti}, {Tartaglia}, {Sand}, {Wyatt},
  {Bostroem}, {Reichart}, {Haislip}, \& {Kouprianov}}]{Valenti2017}
{Valenti}, S., {Tartaglia}, L., {Sand}, D., {et~al.} 2017, The Astronomer's
  Telegram, 10706, 1

\bibitem[{{Van Dyk} {et~al.}(2012{\natexlab{a}}){Van Dyk}, {Cenko},
  {Poznanski}, {Arcavi}, {Gal-Yam}, {Filippenko}, {Silverio}, {Stockton},
  {Cuillandre}, {Marcy}, {Howard}, \& {Isaacson}}]{VanDyk2012b}
{Van Dyk}, S.~D., {Cenko}, S.~B., {Poznanski}, D., {et~al.} 2012{\natexlab{a}},
  \apj, 756, 131

\bibitem[{{Van Dyk} {et~al.}(2012{\natexlab{b}}){Van Dyk}, {Davidge},
  {Elias-Rosa}, {Taubenberger}, {Li}, {Levesque}, {Howerton}, {Pignata},
  {Morrell}, {Hamuy}, \& {Filippenko}}]{VanDyk2012a}
{Van Dyk}, S.~D., {Davidge}, T.~J., {Elias-Rosa}, N., {et~al.}
  2012{\natexlab{b}}, \aj, 143, 19

\bibitem[{{Vink{\'o}} {et~al.}(2006){Vink{\'o}}, {Tak{\'a}ts}, {S{\'a}rneczky},
  {Szab{\'o}}, {M{\'e}sz{\'a}ros}, {Csorv{\'a}si}, {Szalai}, {G{\'a}sp{\'a}r},
  {P{\'a}l}, {Csizmadia}, {K{\'o}sp{\'a}l}, {R{\'a}cz}, {Kun}, {Cs{\'a}k},
  {F{\"u}r{\'e}sz}, {DeBond}, {Grunhut}, {Thomson}, {Mochnacki}, \&
  {Koktay}}]{Vinko2006}
{Vink{\'o}}, J., {Tak{\'a}ts}, K., {S{\'a}rneczky}, K., {et~al.} 2006, \mnras,
  369, 1780

\bibitem[{{Wang} {et~al.}(2001){Wang}, {Howell}, {H{\"o}flich}, \&
  {Wheeler}}]{Wang2001}
{Wang}, L., {Howell}, D.~A., {H{\"o}flich}, P., \& {Wheeler}, J.~C. 2001, \apj,
  550, 1030

\bibitem[{{Wang} \& {Wheeler}(2008)}]{WangWheeler2008}
{Wang}, L. \& {Wheeler}, J.~C. 2008, \araa, 46, 433

\bibitem[{{Wang} {et~al.}(1997){Wang}, {Wheeler}, \& {H{\"o}flich}}]{Wang1997}
{Wang}, L., {Wheeler}, J.~C., \& {H{\"o}flich}, P. 1997, \apjl, 476, L27

\bibitem[{{Whittet} {et~al.}(1992){Whittet}, {Martin}, {Hough}, {Rouse},
  {Bailey}, \& {Axon}}]{Whittet1992}
{Whittet}, D.~C.~B., {Martin}, P.~G., {Hough}, J.~H., {et~al.} 1992, \apj, 386,
  562

\bibitem[{{Zelaya} {et~al.}(2017){Zelaya}, {Clocchiatti}, {Baade},
  {H{\"o}flich}, {Maund}, {Patat}, {Quinn}, {Reilly}, {Wang}, {Wheeler},
  {F{\"o}rster}, \& {Gonz{\'a}lez-Gait{\'a}n}}]{Zelaya2017}
{Zelaya}, P., {Clocchiatti}, A., {Baade}, D., {et~al.} 2017, \apj, 836, 88

\bibitem[{{Zgirski} {et~al.}(2017){Zgirski}, {Gieren}, {Pietrzy{\'n}ski},
  {Karczmarek}, {Gorski}, {Wielgorski}, {Narloch}, {Graczyk}, {Kudritzki}, \&
  {Bresolin}}]{Zgirski2017}
{Zgirski}, B., {Gieren}, W., {Pietrzy{\'n}ski}, G., {et~al.} 2017, \apj, 847,
  88

\end{thebibliography}





\appendix

\section{Observation logs}
\label{app:obs_log}
The observation logs for the spectropolarimetric data are shown here.

\begin{table}
      \caption[]{Log of the observations of SN 2017gmr. The observations were made under the program ID 099.D-0543.
      }
      \label{tab:log_17gmr}
    $\displaystyle
         \begin{array}{lccccccc}
            \hline
            \noalign{\smallskip}
            \rm{Date} & \rm{MJD} & \rm{Airmass} & \rm{Exp. \;time}\\
            (\rm{UT}) & (\rm{days}) & (\rm{average}) & (\rm{s})\\
            \noalign{\smallskip}
            \hline\hline
            \noalign{\smallskip}  
            2017-10-19.60 & 58045.60 & 1.2 & 4 \times 3110\\
            \noalign{\smallskip} \hline \noalign{\smallskip}
            2017-11-10.24 & 58067.24 & 1.2 & 4 \times 660\\
            \noalign{\smallskip} \hline \noalign{\smallskip}
            2017-12-12.86 & 58099.86 & 1.2 & 4 \times 2640\\
            \noalign{\smallskip} \hline \noalign{\smallskip}
            2017-12-21.76 & 58108.76 & 1.2 & 4 \times 3960\\
            \noalign{\smallskip} \hline \noalign{\smallskip}
            2018-01-16.93 & 58134.93 & 1.3 & 4 \times 4620\\
            \noalign{\smallskip}
            \hline
         \end{array}
         $
\end{table}

\begin{table}
      \caption[]{Log of the observations of SN 2017ahn. The observations were made under the program ID 099.D-0543.}
      \label{tab:log_17ahn}
    $\displaystyle
         \begin{array}{lccccccc}
            \hline
            \noalign{\smallskip}
            \rm{Date} & \rm{MJD} & \rm{Airmass} & \rm{Exp. \;time}\\
            (\rm{UT}) & (\rm{days}) & (\rm{average}) & (\rm{s})\\
            \noalign{\smallskip}
            \hline\hline
            \noalign{\smallskip}  
            2017-03-24.75 & 57836.75 & 1.1 & 4 \times 2076\\
            \noalign{\smallskip} \hline \noalign{\smallskip}
            2017-03-31.38 & 57843.38 & 1.2 & 4 \times 2076\\
            \noalign{\smallskip} \hline \noalign{\smallskip}
            2017-04-24.06 & 57867.06 & 1.1 & 4 \times 2076\\
            \noalign{\smallskip} \hline \noalign{\smallskip}
            2017-04-28.73 & 57871.73 & 1.2 & 4 \times 2076\\
            \noalign{\smallskip}
            \hline
         \end{array}
         $
\end{table}

\begin{table}
      \caption[]{Log of the observations of SN 2013ej. The observations were made under the program ID 091.D-0401(A).}
      \label{tab:log_13ej}
    $\displaystyle
         \begin{array}{lccccccc}
            \hline
            \noalign{\smallskip}
            \rm{Date} & \rm{MJD} & \rm{Airmass} & \rm{Exp. \;time}\\
            (\rm{UT}) & (\rm{days}) & (\rm{average}) & (\rm{s})\\
            \noalign{\smallskip}
            \hline\hline
            \noalign{\smallskip}
            2013-08-01.38 & 56505.38 & 1.4 & 4 \times 720\\
            \noalign{\smallskip} \hline \noalign{\smallskip}
            2013-08-27.28 & 56531.28 & 1.5 & 4 \times 900\\
            \noalign{\smallskip} \hline \noalign{\smallskip}
            2013-09-17.21 & 56552.21 & 1.5 & 4 \times 920\\
            \noalign{\smallskip} \hline \noalign{\smallskip}
            2013-09-29.24 & 56564.24 & 1.4 & 4 \times 1480\\
            \noalign{\smallskip} \hline \noalign{\smallskip}
            2013-10-29.23 & 56594.23 & 1.4 & 4 \times 1800\\
            \noalign{\smallskip} \hline \noalign{\smallskip}
            2013-12-04.56 & 56630.56 & 1.4 & 4 \times 3650\\
            \noalign{\smallskip} \hline \noalign{\smallskip}
            2014-01-09.38 & 56666.38 & 1.5 & 4 \times 1800\\
            \noalign{\smallskip}
            \hline
         \end{array}
         $
\end{table}

\begin{table}
      \caption[]{Log of the observations of SN 2012ec. The observations were made under the program ID 089.D-0515(A).}
      \label{tab:log_12ec}
    $\displaystyle
         \begin{array}{lccccccc}
            \hline
            \noalign{\smallskip}
            \rm{Date} & \rm{MJD} & \rm{Airmass} & \rm{Exp. \;time}\\
            (\rm{UT}) & (\rm{days}) & (\rm{average}) & (\rm{s})\\
            \noalign{\smallskip}
            \hline\hline
            \noalign{\smallskip}      
            2012-09-08.28 & 56178.28 & 1.1 & 4 \times 1200\\
            \noalign{\smallskip} \hline \noalign{\smallskip}
            2012-09-25.21 & 56195.21 & 1.1 & 4 \times 1200\\
            \noalign{\smallskip}
            \hline
         \end{array}
         $
\end{table}

\begin{table}
      \caption[]{Log of the observations of SN 2012dh. The observations were made under the program ID 089.D-0515(A).}
      \label{tab:log_12dh}
        $\displaystyle
         \begin{array}{lccccccc}
            \hline
            \noalign{\smallskip}
            \rm{Date} & \rm{MJD} & \rm{Airmass} & \rm{Exp. \;time}\\
            (\rm{UT}) & (\rm{days}) & (\rm{average}) & (\rm{s})\\
            \noalign{\smallskip}
            \hline\hline
            \noalign{\smallskip}      
            2012-07-18.03 & 56126.03 & 1.3 & 4 \times 1800\\
            \noalign{\smallskip} \hline \noalign{\smallskip}
            2012-08-11.98 & 56150.98 & 1.8 & 4 \times 1800\\
            \noalign{\smallskip} \hline \noalign{\smallskip}
            2012-08-24.99 & 56163.99 & 1.9 & 4 \times 900\\
            \noalign{\smallskip} \hline \noalign{\smallskip}
            2012-08-27.01 & 56166.01 & 2.2 & 4 \times 440\\
            \noalign{\smallskip}
            \hline
         \end{array}
         $
\end{table}

\begin{table}
      \caption[]{Log of the observations of SN 2012aw. The observations were made under the program ID 089.D-0515(A).}
      \label{tab:log_12aw}
        $\displaystyle
         \begin{array}{lccccccc}
            \hline
            \noalign{\smallskip}
            \rm{Date} & \rm{MJD} & \rm{Airmass} & \rm{Exp. \;time}\\
            (\rm{UT}) & (\rm{days}) & (\rm{average}) & (\rm{s})\\
            \noalign{\smallskip}
            \hline\hline
            \noalign{\smallskip}      
            2012-04-01.19 & 56018.19 & 1.5 & 4 \times 400\\
            \noalign{\smallskip} \hline \noalign{\smallskip}
            2012-05-01.04 & 56048.04 & 1.2 & 4 \times 200\\
            \noalign{\smallskip} \hline \noalign{\smallskip}
            2012-05-18.07 & 56065.07 & 1.5 & 4 \times 1620\\
            \noalign{\smallskip} \hline \noalign{\smallskip}
            2012-05-27.01 & 56074.01 & 1.6 & 4 \times 1800\\
            \noalign{\smallskip} \hline \noalign{\smallskip}
            2012-06-16.03 & 56094.03 & 1.9 & 4 \times 1215\\
            \noalign{\smallskip} \hline \noalign{\smallskip}
            2012-07-02.01 & 56110.01 & 2.1 & 4 \times 800\\ 
            \noalign{\smallskip} \hline \noalign{\smallskip}
            2012-07-15.98 & 56123.98 & 2.5 & 4 \times 300\\
            \noalign{\smallskip}
            \hline
         \end{array}
         $
\end{table}

\begin{table}
      \caption[]{Log of the observations of SN 2010hv. The observations were made under the program ID 085.D-0391.}
      \label{tab:log_10hv}
        $\displaystyle
         \begin{array}{lccccccc}
            \hline
            \noalign{\smallskip}
            \rm{Date} & \rm{MJD} & \rm{Airmass} & \rm{Exp. \;time}\\
            (\rm{UT}) & (\rm{days}) & (\rm{average}) & (\rm{s})\\
            \noalign{\smallskip}
            \hline\hline
            \noalign{\smallskip}      
            2010-09-21.15 & 55460.15 & 1.0 & 4 \times 900\\
            \noalign{\smallskip} \hline \noalign{\smallskip}
            2010-09-24.14 & 55463.14 & 1.0 & 4 \times 900\\
            \noalign{\smallskip} \hline \noalign{\smallskip}
            2010-10-15.07 & 55484.07 & 1.0 & 4 \times 900\\
            \noalign{\smallskip} \hline \noalign{\smallskip}
            2010-11-29.56 & 55529.56 & 1.2 & 4 \times 3600\\
            \noalign{\smallskip}
            \hline
         \end{array}
         $
\end{table}

\begin{table}
      \caption[]{Log of the observations of SN 2010co. The observations were made under the program ID 085.D-0391.}
      \label{tab:log_10co}
        $\displaystyle
         \begin{array}{lccccccc}
            \hline
            \noalign{\smallskip}
            \rm{Date} & \rm{MJD} & \rm{Airmass} & \rm{Exp. \;time}\\
            (\rm{UT}) & (\rm{days}) & (\rm{average}) & (\rm{s}) \\
            \noalign{\smallskip}
            \hline\hline
            \noalign{\smallskip}      
            2010-06-05.73 & 55352.73 & 1.3 & 4 \times 1800\\
            \noalign{\smallskip} \hline \noalign{\smallskip}
            2010-07-08.21 & 55385.21 & 1.2 & 4 \times 900\\
            \noalign{\smallskip} \hline \noalign{\smallskip}
            (2010-08-01.21 & 55409.21 & 1.3 & 4 \times 5400)\\
            \noalign{\smallskip}
            \hline
         \end{array}
         $
\end{table}

\begin{table}
      \caption[]{Log of the observations of SN 2008bk. The observations were made under the program IDs 081.D-0128(A) and 082.D-0151(A).}
      \label{tab:log_08bk}
        $\displaystyle
         \begin{array}{lccccccc}
            \hline
            \noalign{\smallskip}
            \rm{Date} & \rm{MJD} & \rm{Airmass} & \rm{Exp. \;time}\\
            (\rm{UT}) & (\rm{days}) & (\rm{average}) & (\rm{s})\\
            \noalign{\smallskip}
            \hline\hline
            \noalign{\smallskip}      
            2008-06-02.36 & 54619.36 & 1.2 & 4 \times 750\\
            \noalign{\smallskip} \hline \noalign{\smallskip}
            2008-07-01.31 & 54648.31 & 1.1 & 4 \times 435\\
            \noalign{\smallskip} \hline \noalign{\smallskip}
            2008-07-24.29 & 54671.29 & 1.0 & 4 \times 1800\\
            \noalign{\smallskip} \hline \noalign{\smallskip}
            2008-07-30.28 & 54677.28 & 1.0 & 4 \times 1800\\
            \noalign{\smallskip} \hline \noalign{\smallskip}
            2008-08-06.25 & 54684.25 & 1.1 & 4 \times 2230\\
            \noalign{\smallskip} \hline \noalign{\smallskip}
            2008-09-28.10 & 54737.10 & 1.1 & 4 \times 2330\\
            \noalign{\smallskip} \hline \noalign{\smallskip}
            2008-11-19.03 & 54789.03 & 1.2 & 4 \times 6450\\
            \noalign{\smallskip} \hline \noalign{\smallskip}
            2008-12-20.04 & 54820.04 & 1.5 & 4 \times 4000\\
            \noalign{\smallskip} \hline \noalign{\smallskip}
            2009-01-01.04 & 54832.04 & 1.9 & 4 \times 5770\\
            \noalign{\smallskip}
            \hline
         \end{array}
         $
\end{table}

\begin{table}
      \caption[]{Log of the observations of SN 2001du. The observations were made under the program ID 67.D-0517(A).}
      \label{tab:log_01du}
        $\displaystyle
         \begin{array}{lccccccc}
            \hline
            \noalign{\smallskip}
            \rm{Date} & \rm{MJD} & \rm{Airmass} & \rm{Exp. \;time}\\
            (\rm{UT}) & (\rm{days}) & (\rm{average}) & (\rm{s})\\
            \noalign{\smallskip}
            \hline\hline
            \noalign{\smallskip}      
            2001-08-30.32 & 52151.32 & 1.1 & 4 \times 1200\\
            \noalign{\smallskip} \hline \noalign{\smallskip}
            2001-09-13.36 & 52165.36 & 1.0 & 4 \times 1200\\
            \noalign{\smallskip}
            \hline
         \end{array}
         $
\end{table}

\begin{table}
      \caption[]{Log of the observations of SN 2001dh. The observations were made under the program ID 67.D-0517(A).}
      \label{tab:log_01dh}
        $\displaystyle
         \begin{array}{lccccccc}
            \hline
            \noalign{\smallskip}
            \rm{Date} & \rm{MJD} & \rm{Airmass} & \rm{Exp. \;time}\\
            (\rm{UT}) & (\rm{days}) & (\rm{average}) & (\rm{s})\\
            \noalign{\smallskip}
            \hline\hline
            \noalign{\smallskip}      
            2001-08-10.18 & 52131.18 & 1.1 & 4 \times 1200\\
            \noalign{\smallskip} \hline \noalign{\smallskip}
            2001-08-19.16 & 52140.16 & 1.1 & 4 \times 1200\\
            \noalign{\smallskip} \hline \noalign{\smallskip}
            2001-08-29.05 & 52150.05 & 1.0 & 4 \times 1200\\
            \noalign{\smallskip} \hline \noalign{\smallskip}
            2001-09-11.09 & 52163.09 & 1.0 & 4 \times 600\\
            \noalign{\smallskip}
            \hline
         \end{array}
         $
\end{table}

\section{SN~2007aa} 
\label{app:07aa}

\subsection{Photometry}
\label{app:B1}

\begin{figure*}
    \includegraphics[width=2\columnwidth]{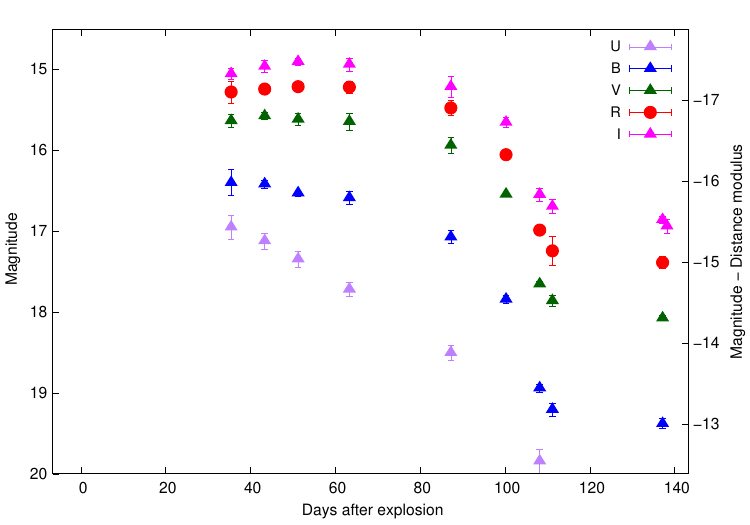}
    \caption{Photometry of SN~2007aa. The assumed explosion date is 54126.7 MJD.
    }
    \label{fig:app_LC_07aa}
\end{figure*}

We obtained $U\!BV\!RI$-band images of SN~2007aa with the CAFOS (Calar Alto Faint Object Spectrograph) instrument, the visual imager and low-resolution spectrograph at the Calar Alto 2.2m Telescope.  All frames were de-biased and flatfield-corrected using standard techniques with IRAF.  \citet{Bessell1990} magnitudes of local field stars were derived from their Sloan $ugri$ magnitudes reported in the SDSS catalogue using the transformation equations by \citet{Jester2005}.  Instrumental SN magnitudes were measured through PSF photometry with SNOOPY\footnote{SNOOPY is a collection of scripts, originally developed by F. Patat and later implemented in IRAF by E. Cappellaro. It is optimised for PSF-fitting photometry of point sources superimposed on a structured background.}, subtracting a local fit to the underlying host-galaxy light. The instrumental SN magnitudes were then calibrated to the Bessell (1990) system with zero points obtained from the local sequence stars and $S$-corrections \citep{Stritzinger2002} calculated with the help of the spectra described in Section~\ref{app:B2}.  Photometric uncertainties were estimated as the quadratic sum of PSF-fitting uncertainties, uncertainties in the PSF-to-aperture correction, uncertainties in the photometric zero points, and uncertainties in the background fit estimated through an artificial-star experiment.  In reality, these contributions may not be fully independent, hence our errors might be over-estimated in some cases. The resulting $S$-corrected photometry along with the derived uncertainties is provided in Figure~\ref{fig:app_LC_07aa} and Table~\ref{tab:photo_07aa}.


\subsection{Spectroscopy}
\label{app:B2}

\begin{figure*}
    \includegraphics[width=2\columnwidth]{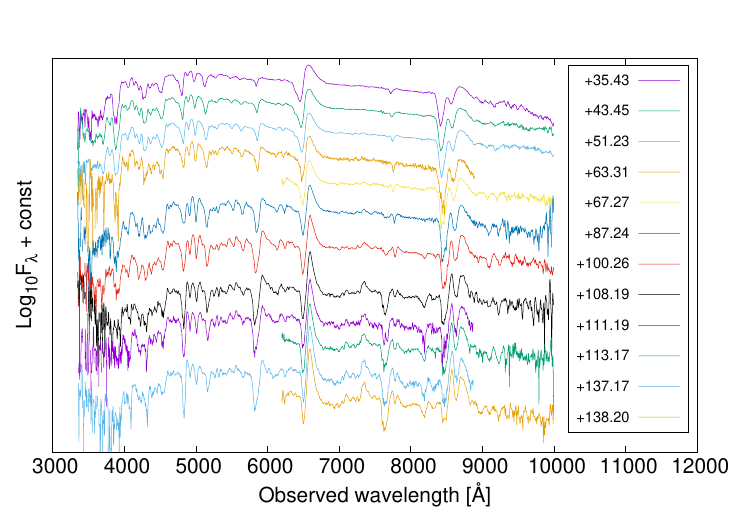}
    \caption{Spectra of SN~2007aa. The days after the assumed explosion date (54126.7 MJD) are shown.}
    \label{fig:app_spec_07aa}
\end{figure*}

We also obtained optical spectroscopy of SN~2007aa with the CAFOS instrument at the Calar Alto 2.2m Telescope, using the b200 and r200 grisms, and a 1.5" slit aligned along the parallactic angle to minimise differential slit losses.  After the usual de-biasing and flatfielding, a variance-weighted extraction of the spectra was performed in the IRAF task APALL. The wavelength solution was established with the help of arc-lamp spectra and checked against night-sky emission lines.  The flux calibration was performed with the help of sensitivity curves derived from standard-star observations. b200 and r200 spectra obtained during the same night were combined to increase the wavelength coverage and the S/N in the overlap region.  All spectra were corrected for telluric absorptions on a best-effort basis.  As a final step, the flux of the spectra was checked against the contemporaneous $BV\!RI$ photometry (Section~\ref{app:B1}) and adjusted with linear functions if necessary. A log of the spectroscopic observations is reported in Table~\ref{tab:spec_log_07aa}.  The spectra are shown in Figure~\ref{fig:app_spec_07aa}.


   \begin{table*}
      \caption[]{Photometry of SN~2007aa}
         \label{tab:photo_07aa}
     $$
         \tiny
         \begin{tabular}{cccccc}
            \hline
            \noalign{\smallskip}
            MJD & $U$ & $B$ & $V$ & $R$ & $I$\\
            \noalign{\smallskip}
            \hline
            \noalign{\smallskip}
            54162.13 & 17.06 (0.15) & 16.49 (0.16) & 15.70 (0.08) & 15.33 (0.14) & 15.09 (0.07)\\
            54170.04 & 17.23 (0.10) & 16.51 (0.05) & 15.64 (0.04) & 15.30 (0.04) & 15.00 (0.07)\\
            54177.91 & 17.46 (0.10) & 16.62 (0.04) & 15.69 (0.07) & 15.27 (0.04) & 14.94 (0.04)\\
            54190.01 & 17.83 (0.09) & 16.68 (0.08) & 15.71 (0.10) & 15.27 (0.07) & 14.97 (0.08)\\
            54213.93 & 18.61 (0.09) & 17.17 (0.08) & 16.01 (0.10) & 15.53 (0.09) & 15.25 (0.13)\\
            54226.95 & -            & 17.94 (0.05) & 16.61 (0.03) & 16.11 (0.05) & 15.69 (0.06)\\
            54234.88 & 19.96 (0.15) & 19.03 (0.05) & 17.72 (0.03) & 17.04 (0.05) & 16.58 (0.08)\\
            54237.87 & -            & 19.30 (0.08) & 17.93 (0.06) & 17.30 (0.18) & 16.73 (0.09)\\
            54263.87 & -            & 19.47 (0.06) & 18.14 (0.03) & 17.44 (0.08) & 16.90 (0.05)\\
            54264.89 & -            & -            & -            & -            & 16.98 (0.09)\\
            \noalign{\smallskip}
            \hline\\
        \end{tabular}
     $$
   \end{table*}

   \begin{table*}
      \caption[]{Log of the spectroscopy of SN~2007aa.}
         \label{tab:spec_log_07aa}
     $$
         \begin{tabular}{cccc}
            \hline
            \noalign{\smallskip}
            Date (UT) & MJD & Telescope & Instrument\\
            \noalign{\smallskip}
            \hline
            \noalign{\smallskip}
            2007 March 03.13 & 54162.13 & Calar Alto 2.2m telescope & CAFOS\\
            2007 March 11.05 & 54170.05 & Calar Alto 2.2m telescope & CAFOS\\
            2007 March 18.93 & 54177.93 & Calar Alto 2.2m telescope & CAFOS\\
            2007 March 31.01 & 54190.01 & Calar Alto 2.2m telescope & CAFOS\\
            2007 April 03.97 & 54193.97 & Calar Alto 2.2m telescope & CAFOS\\
            2007 April 23.94 & 54213.94 & Calar Alto 2.2m telescope & CAFOS\\
            2007 May 06.96 & 54226.96 & Calar Alto 2.2m telescope & CAFOS\\
            2007 May 14.89 & 54234.89 & Calar Alto 2.2m telescope & CAFOS\\
            2007 May 17.89 & 54237.89 & Calar Alto 2.2m telescope & CAFOS\\
            2007 May 19.87 & 54239.87 & Calar Alto 2.2m telescope & CAFOS\\
            2007 June 12.90 & 54263.90 & Calar Alto 2.2m telescope & CAFOS\\
            2007 June 13.90 & 54264.90 & Calar Alto 2.2m telescope & CAFOS\\
            \noalign{\smallskip}
            \hline
        \end{tabular}
     $$
   \end{table*}

\section{Spectroscopic data of the FORS/VLT sample} 
\label{app:spec}

The spectra obtained as the spectropolarimetric observations for the VLT sample are provided here. All the spectra are normalized by their continuum flux using the IRAF onedspec.continuum task with the cubic spline curve and the rejection limit of 2 residual sigma.

\begin{figure}
    \includegraphics[width=\columnwidth]{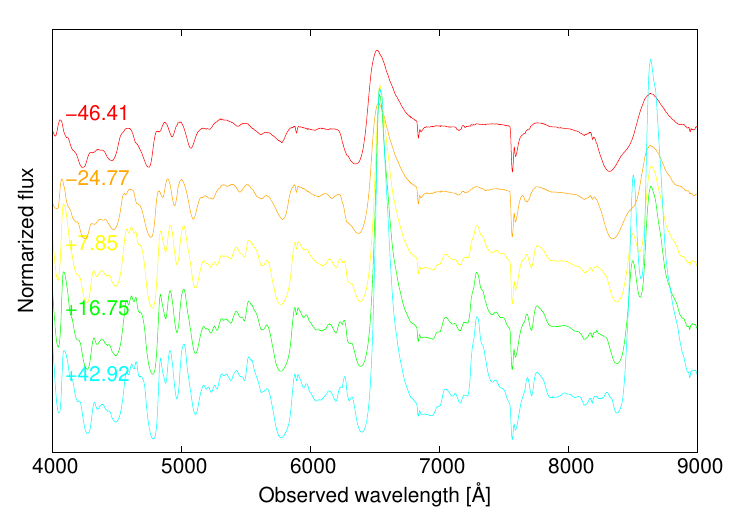}
    \caption{FORS spectrum of SN 2017gmr. The top-left number in each panel shows the phase of the spectrum (days relative to end of the photospheric phase; see Paper~II).}
    \label{fig:app3_17gmr}
\end{figure}

\begin{figure}
    \includegraphics[width=\columnwidth]{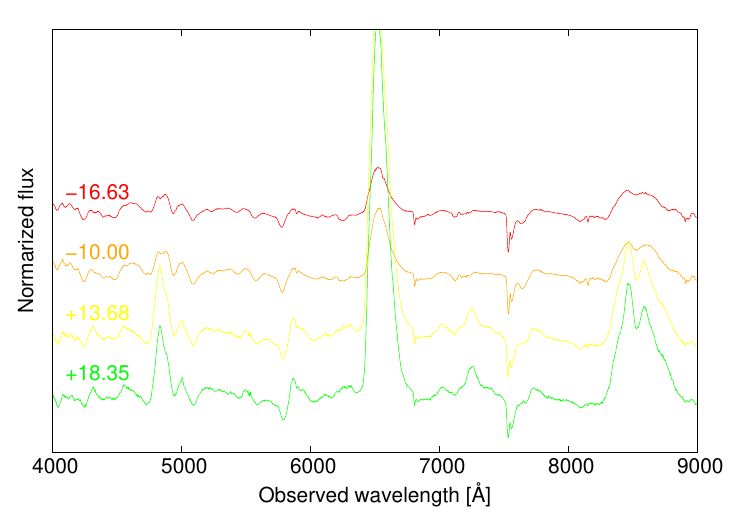}
    \caption{Same as Figure~\ref{fig:app3_17gmr}, but for SN~2017ahn.}
\end{figure}

\begin{figure}
    \includegraphics[width=\columnwidth]{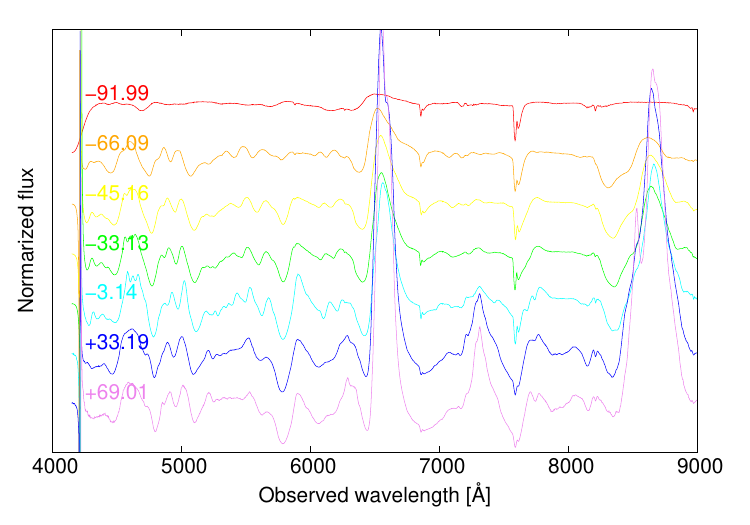}
    \caption{Same as Figure~\ref{fig:app3_17gmr}, but for SN 2013ej.}
\end{figure}

\begin{figure}
    \includegraphics[width=\columnwidth]{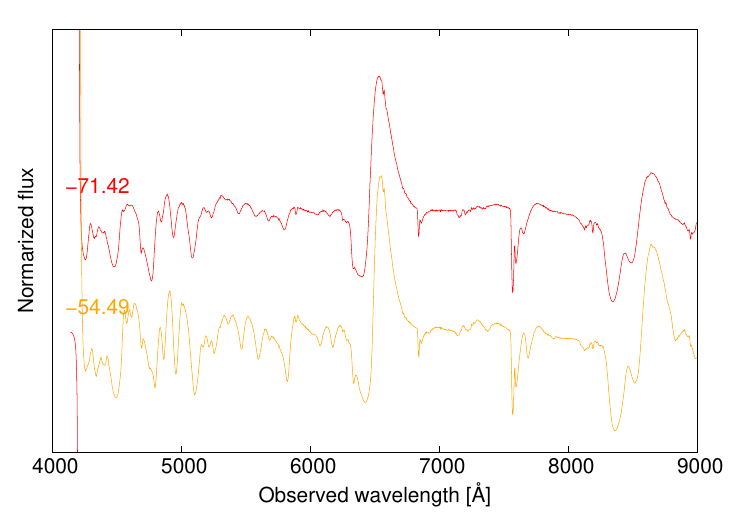}
    \caption{Same as Figure~\ref{fig:app3_17gmr}, but for SN 2012ec.}
\end{figure}

\begin{figure}
\includegraphics[width=\columnwidth]{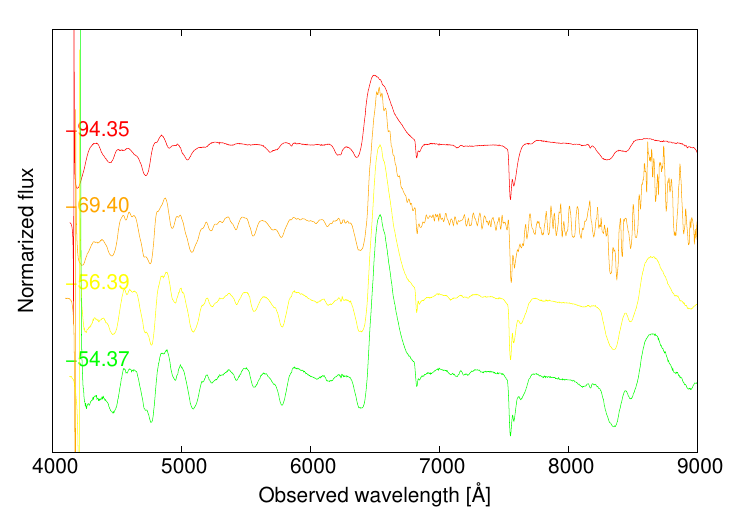}
    \caption{Same as Figure~\ref{fig:app3_17gmr}, but for SN 2012dh.}
\end{figure}

\begin{figure}
    \includegraphics[width=\columnwidth]{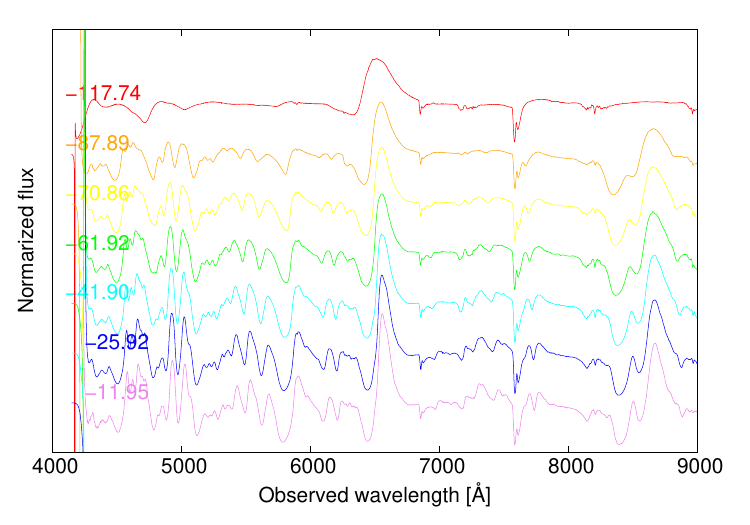}
    \caption{Same as Figure~\ref{fig:app3_17gmr}, but for SN 2012aw.}
\end{figure}

\begin{figure}
    \includegraphics[width=\columnwidth]{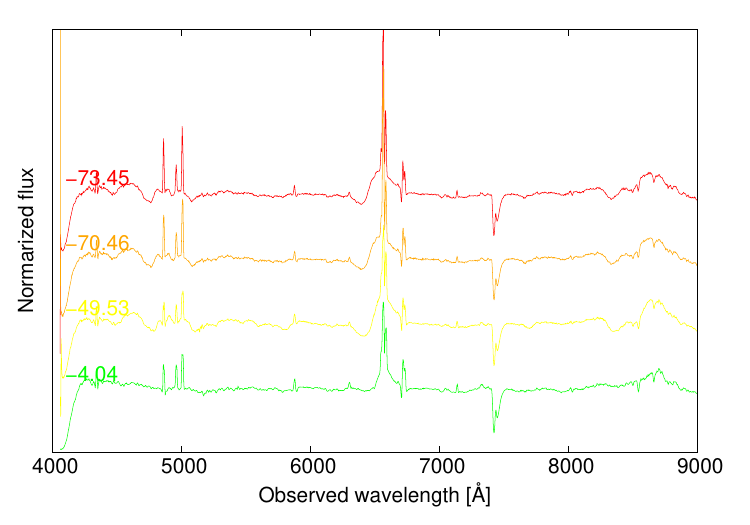}
    \caption{Same as Figure~\ref{fig:app3_17gmr}, but for SN 2010hv.}
\end{figure}

\begin{figure}
    \includegraphics[width=\columnwidth]{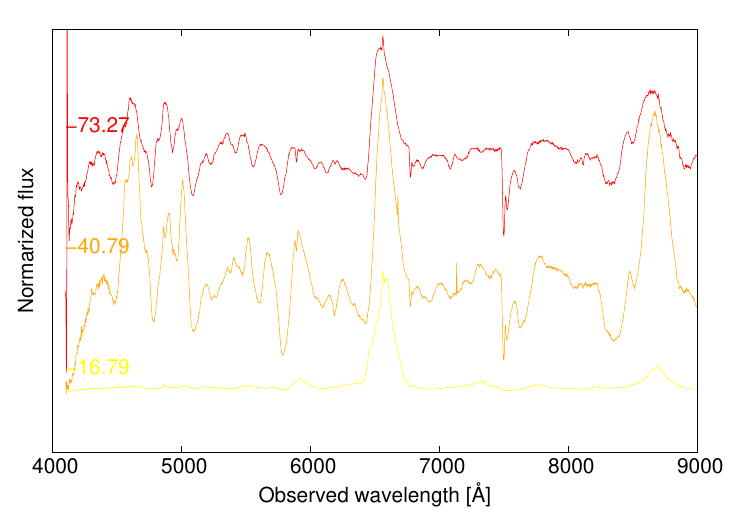}
    \caption{Same as Figure~\ref{fig:app3_17gmr}, but for SN 2010co.}
\end{figure}

\begin{figure}
    \includegraphics[width=\columnwidth]{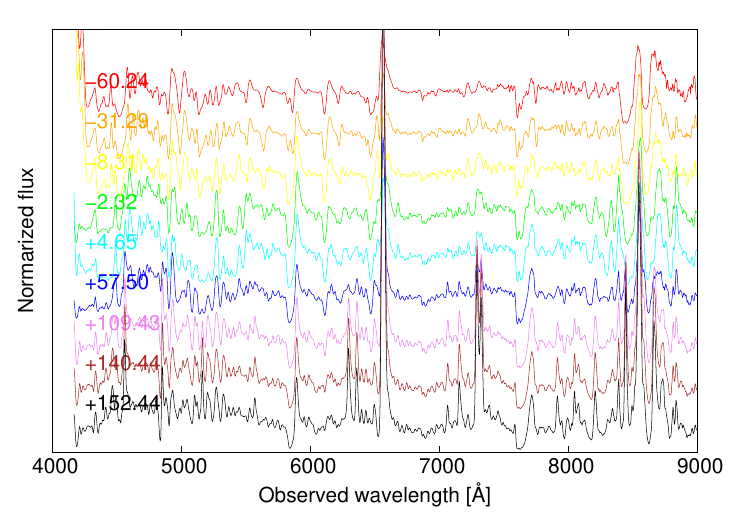}
    \caption{Same as Figure~\ref{fig:app3_17gmr}, but for SN 2008bk.}
\end{figure}

\begin{figure}
    \includegraphics[width=\columnwidth]{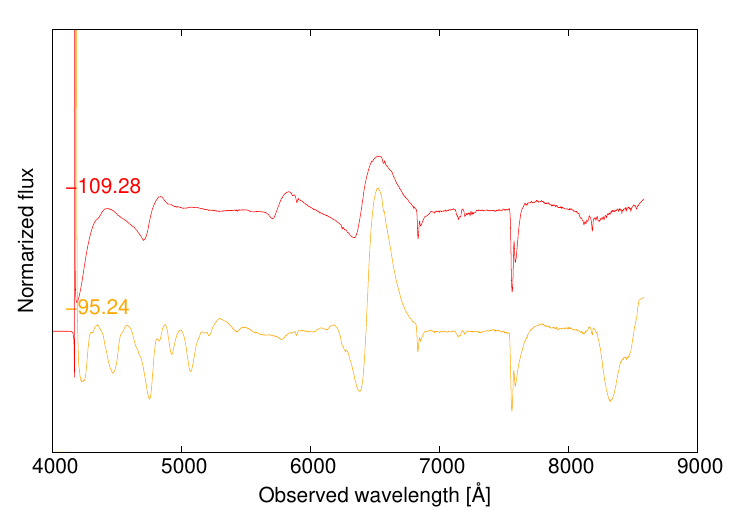}
    \caption{Same as Figure~\ref{fig:app3_17gmr}, but for SN 2001du.}
\end{figure}

\begin{figure}
    \includegraphics[width=\columnwidth]{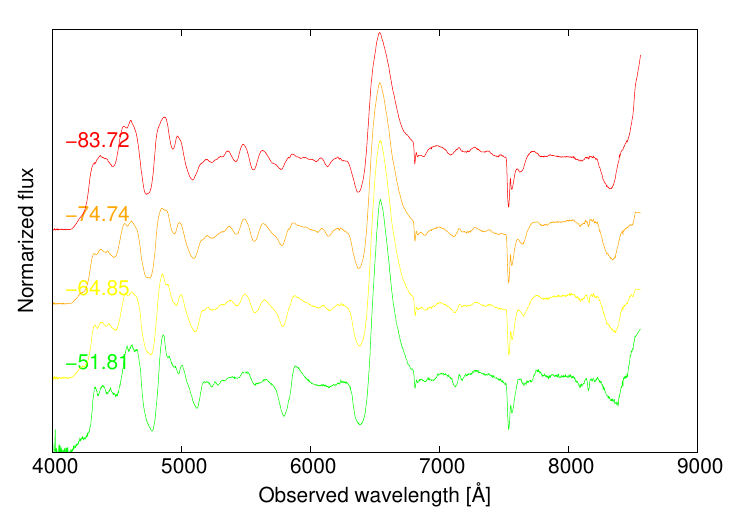}
    \caption{Same as Figure~\ref{fig:app3_17gmr}, but for SN 2001dh.}
\end{figure}

\section{Polarization spectra}
\label{app:pol_spec}
The polarization spectra of all the SNe before ISP subtraction are shown here. In order to increase the signal-to-noise ratio, the spectra have been rebinned to 25-\AA\; bins here.

\begin{figure*}
  \includegraphics[width=2\columnwidth]{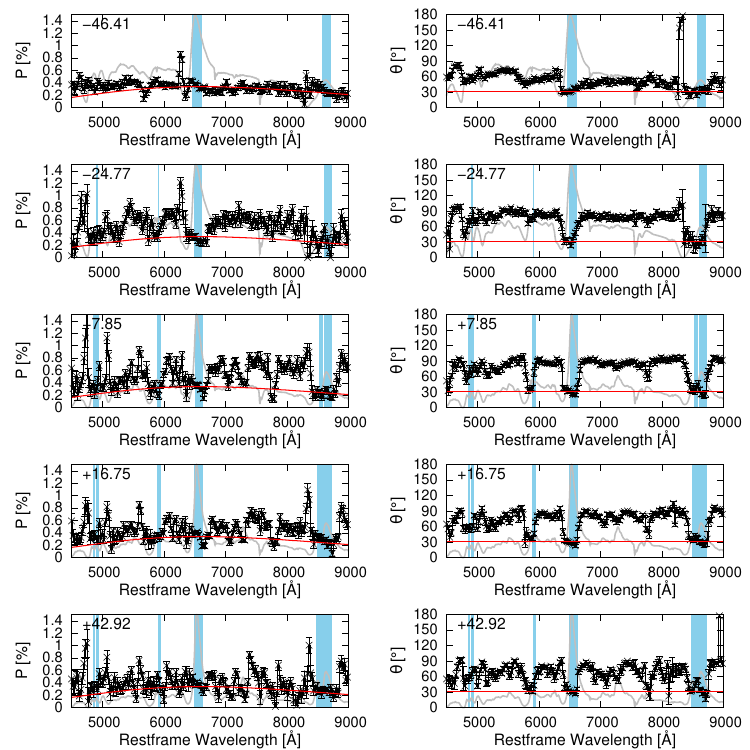}
      \caption{
      Polarization degree (left) and angle (right) of SN~2017gmr before the ISP subtraction at different epochs (increasing from the top to the bottom as labelled). The grey lines in the background of each plot are the unbinned flux spectra at the same epochs. The blue shading shows the adopted wavelength ranges for the emission lines at each epoch. The red lines show the best-fit ISP. The top-left number in each panel shows the phase (days relative to end of the photospheric phase; see Paper~II).
    }
    \label{fig:app_17gmr}
\end{figure*}

\begin{figure*}
  \includegraphics[width=2\columnwidth]{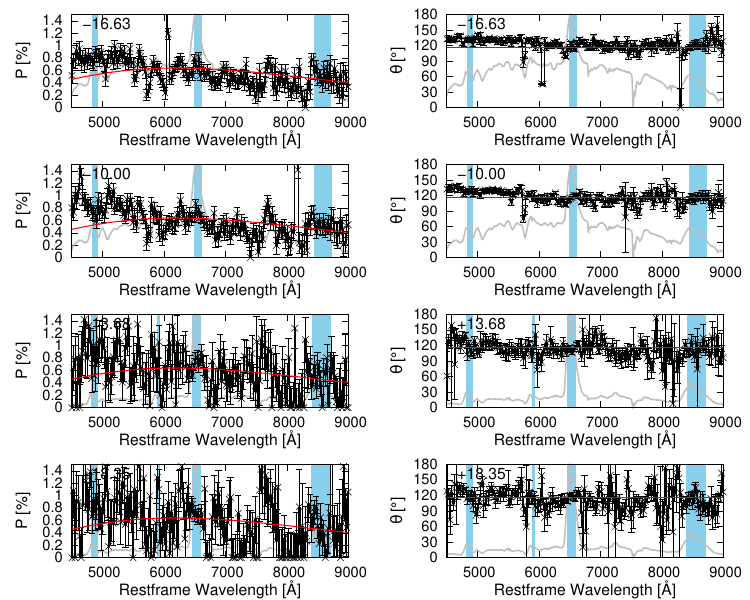}
      \caption{Same as Fig.~\ref{fig:app_17gmr}, but for SN~2017ahn.
    }
    \label{fig:app_17ahn}
\end{figure*}

\begin{figure*}
  \includegraphics[width=2\columnwidth]{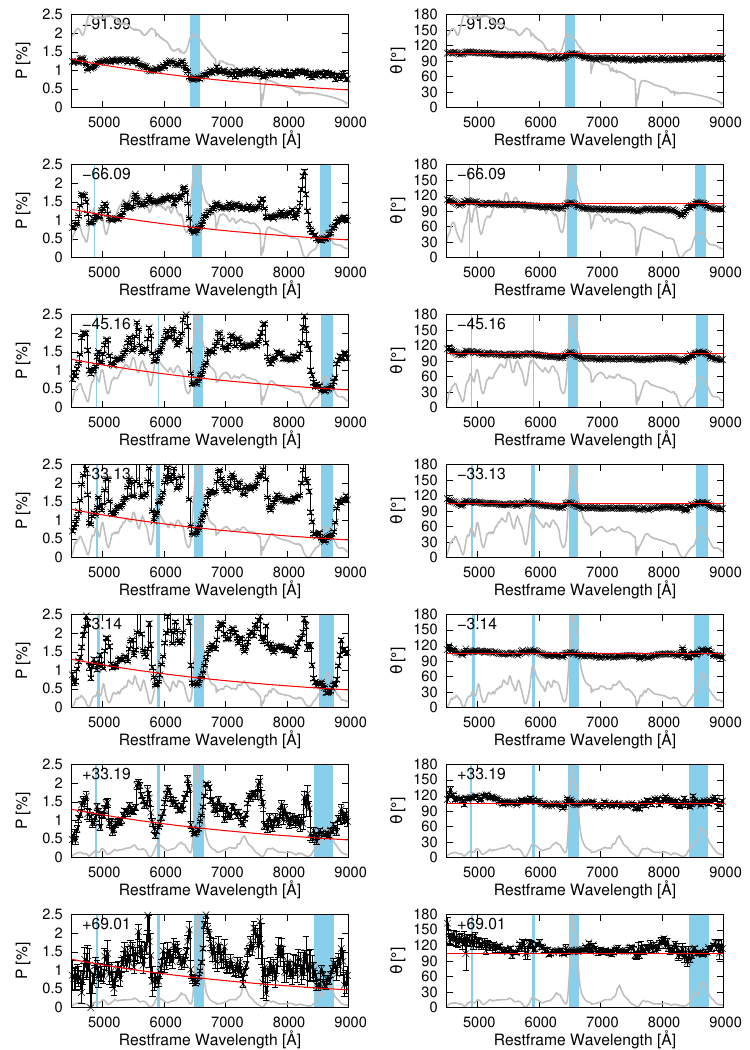}
      \caption{Same as Fig.~\ref{fig:app_17gmr}, but for SN~2013ej.
    }
    \label{fig:app_13ej}
\end{figure*}

\begin{figure*}
  \includegraphics[width=2\columnwidth]{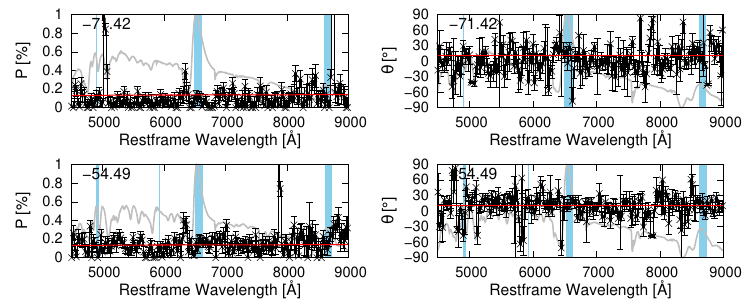}
      \caption{Same as Fig.~\ref{fig:app_17gmr}, but for SN~2012ec.
    }
    \label{fig:app_12ec}
\end{figure*}

\begin{figure*}
  \includegraphics[width=2\columnwidth]{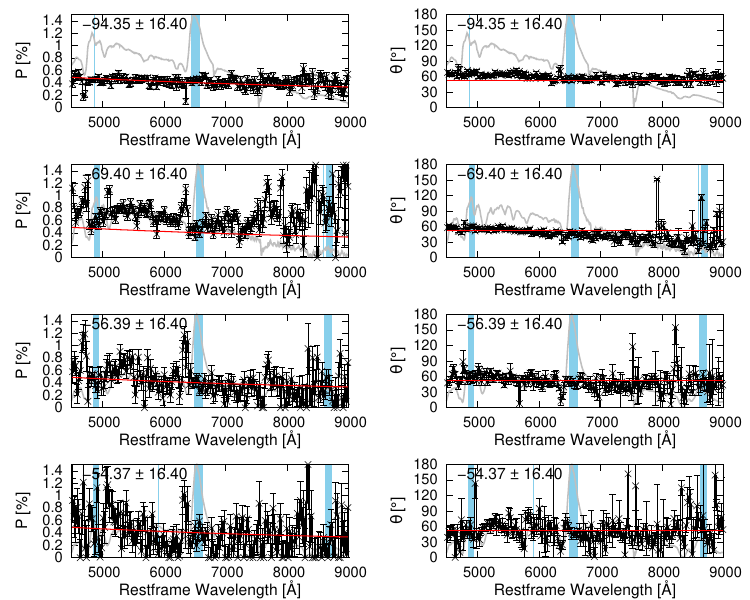}
      \caption{Same as Fig.~\ref{fig:app_17gmr}, but for SN~2012dh.
    }
    \label{fig:app_12dh}
\end{figure*}

\begin{figure*}
  \includegraphics[width=2\columnwidth]{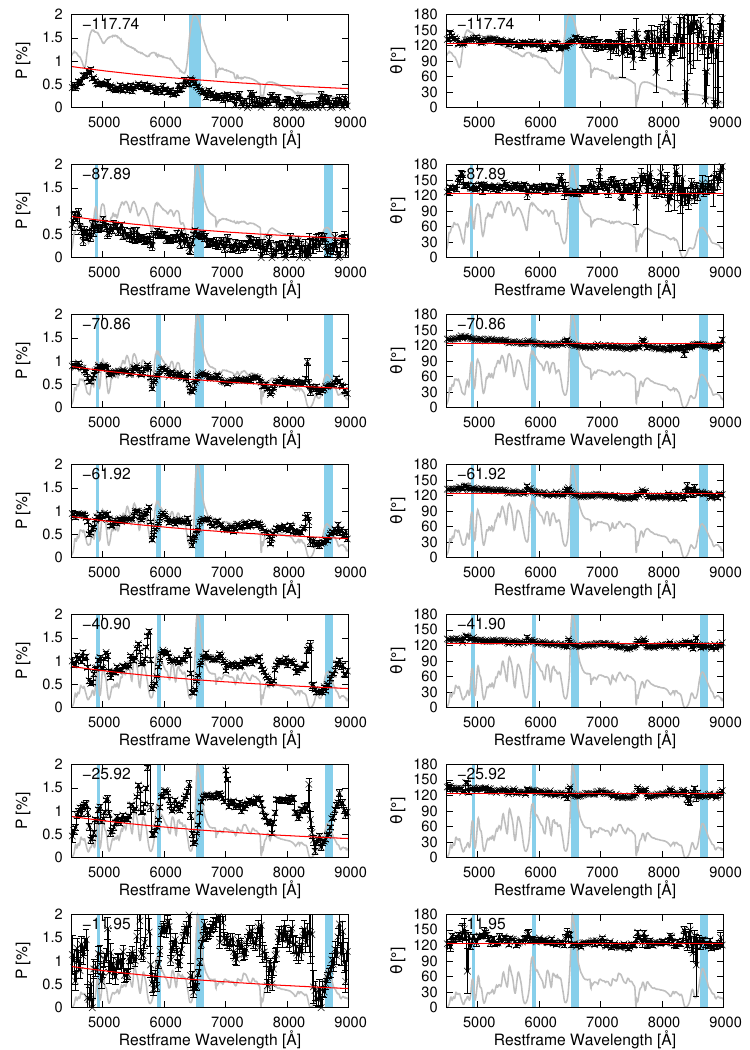}
      \caption{Same as Fig.~\ref{fig:app_17gmr}, but for SN~2012aw.
    }
    \label{fig:app_12aw}
\end{figure*}

\begin{figure*}
  \includegraphics[width=2\columnwidth]{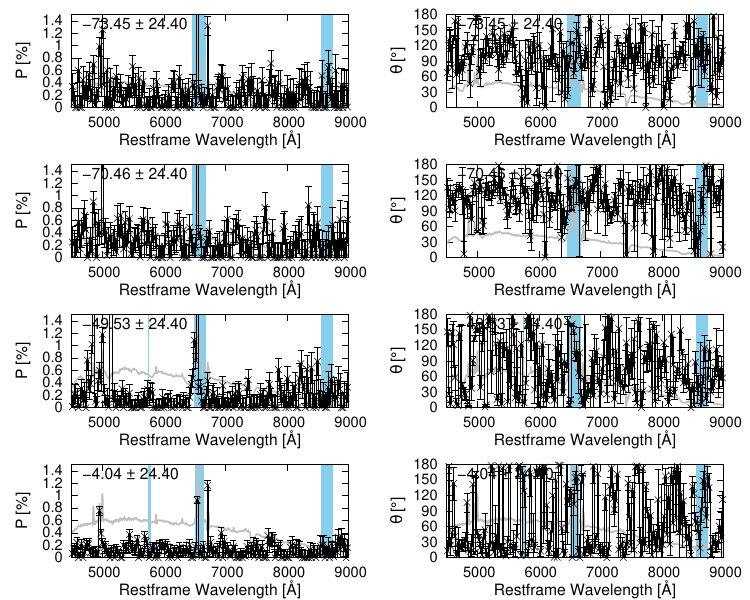}
      \caption{Same as Fig.~\ref{fig:app_17gmr}, but for SN~2010hv.
    }
    \label{fig:app_10hv}
\end{figure*}

\begin{figure*}
  \includegraphics[width=2\columnwidth]{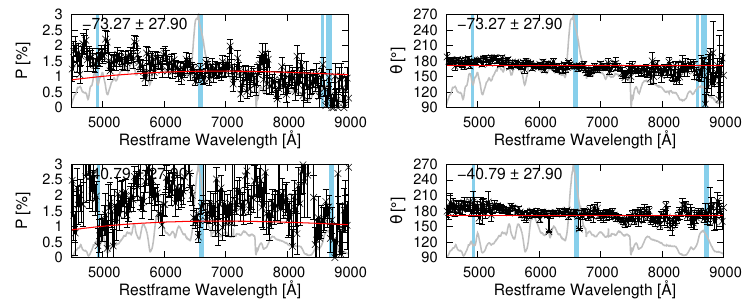}
      \caption{Same as Fig.~\ref{fig:app_17gmr}, but for SN~2010co.
    }
    \label{fig:app_10co}
\end{figure*}

\begin{figure*}
  \includegraphics[width=1.5\columnwidth]{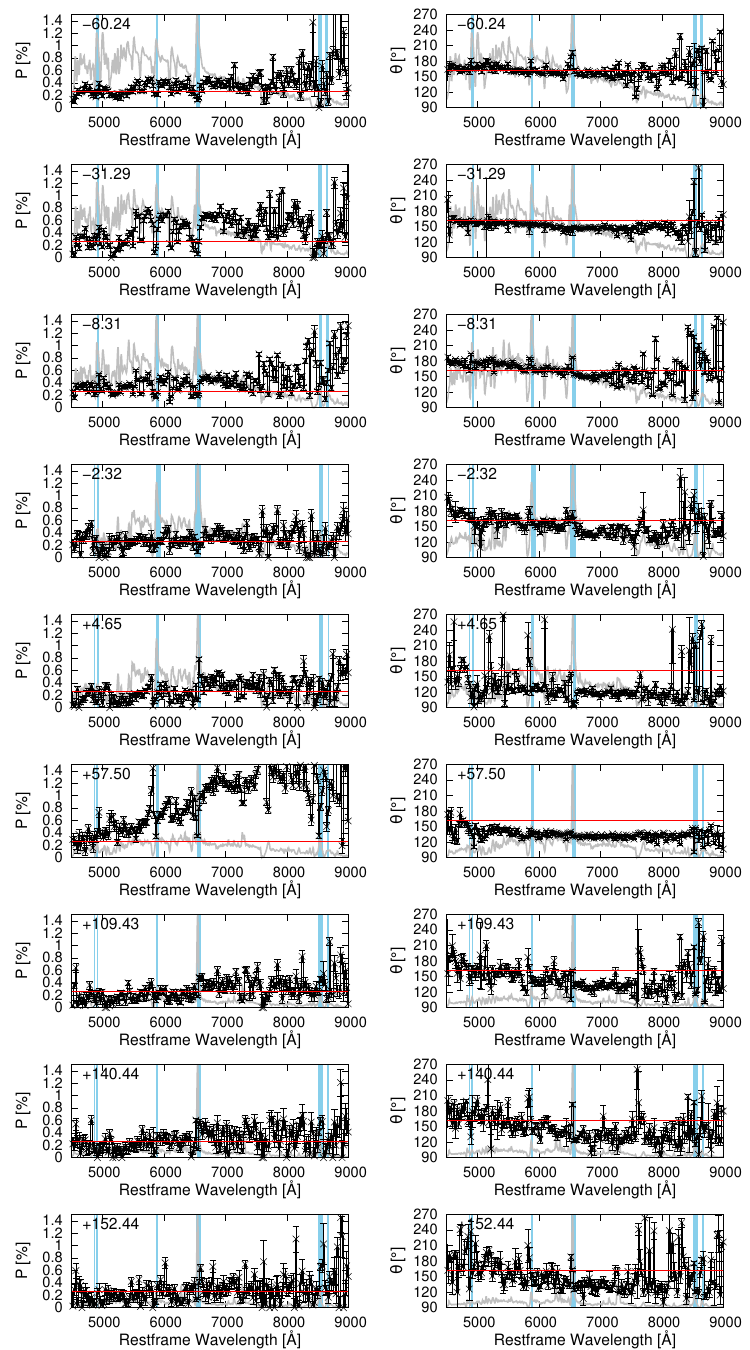}
      \caption{Same as Fig.~\ref{fig:app_17gmr}, but for SN~2008bk.
    }
    \label{fig:app_08bk}
\end{figure*}

\begin{figure*}
  \includegraphics[width=2\columnwidth]{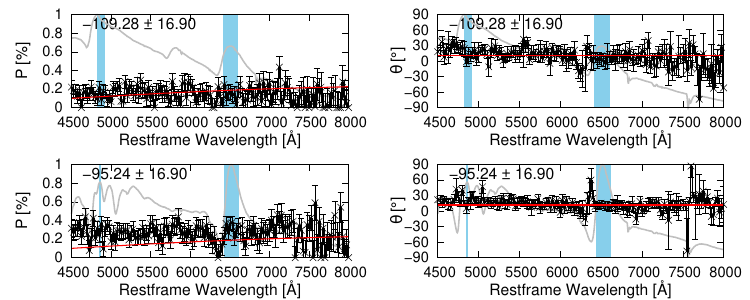}
      \caption{Same as Fig.~\ref{fig:app_17gmr}, but for SN~2001du.
    }
    \label{fig:app_01du}
\end{figure*}

\begin{figure*}
  \includegraphics[width=2\columnwidth]{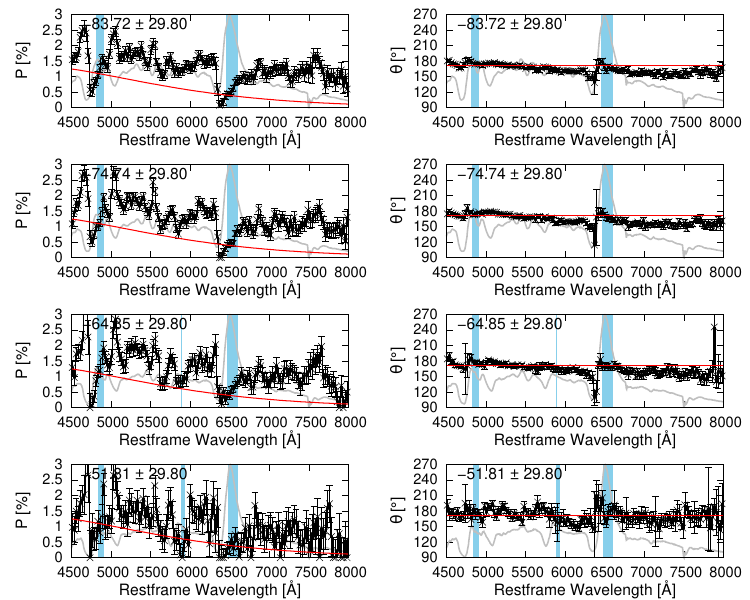}
      \caption{Same as Fig.~\ref{fig:app_17gmr}, but for SN~2001dh.
    }
    \label{fig:app_01dh}
\end{figure*}

\section{Extracted ISP components}
\label{app:ISP_spec}
The extracted ISP components for all the SNe are shown here.

\begin{figure*}
  \includegraphics[width=2\columnwidth]{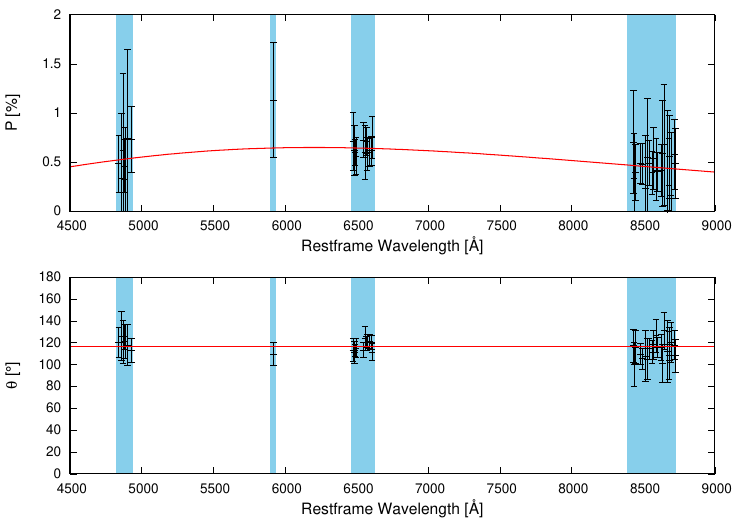}
      \caption{Polarization degree $P$ and angle $\theta$ selected for the ISP in SN~2017ahn. The red lines and the blue shading are the same as in Fig.~\ref{fig:ISP_17gmr}.
    }
    \label{fig:app_ISP_17ahn}
\end{figure*}

\begin{figure*}
  \includegraphics[width=2\columnwidth]{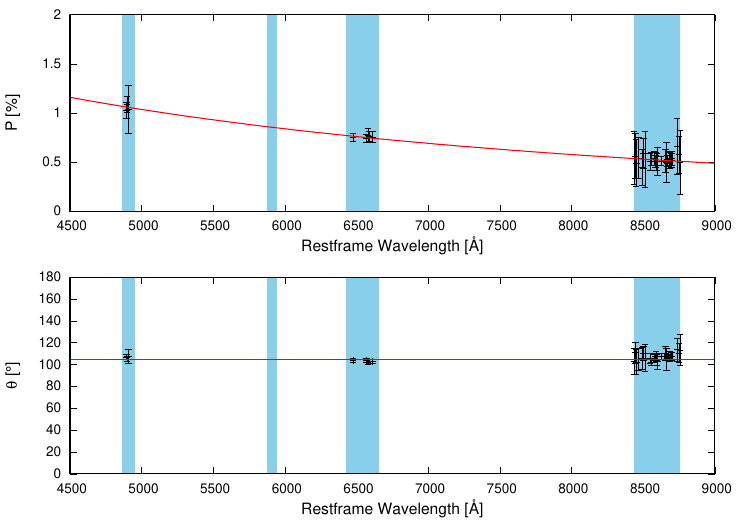}
      \caption{Same as Fig.~\ref{fig:app_ISP_17ahn}, but for SN~2013ej.
    }
    \label{fig:app_ISP_13ej}
\end{figure*}

\begin{figure*}
  \includegraphics[width=2\columnwidth]{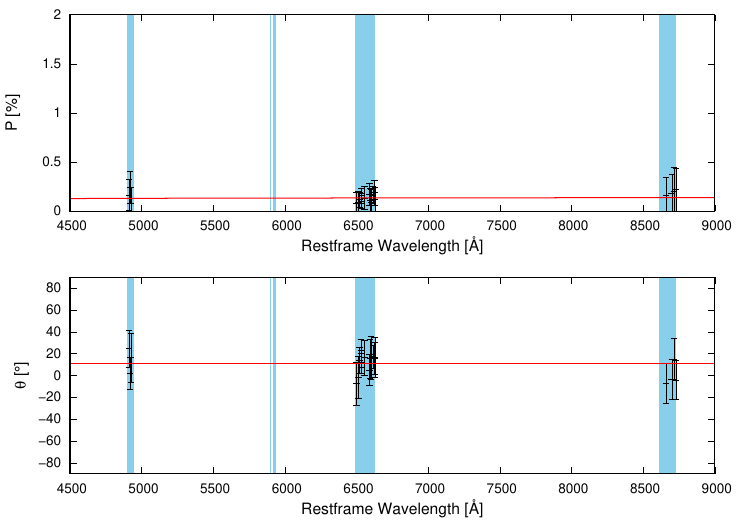}
      \caption{Same as Fig.~\ref{fig:app_ISP_17ahn}, but for SN~2012ec.
    }
    \label{fig:app_ISP_12ec}
\end{figure*}

\begin{figure*}
  \includegraphics[width=2\columnwidth]{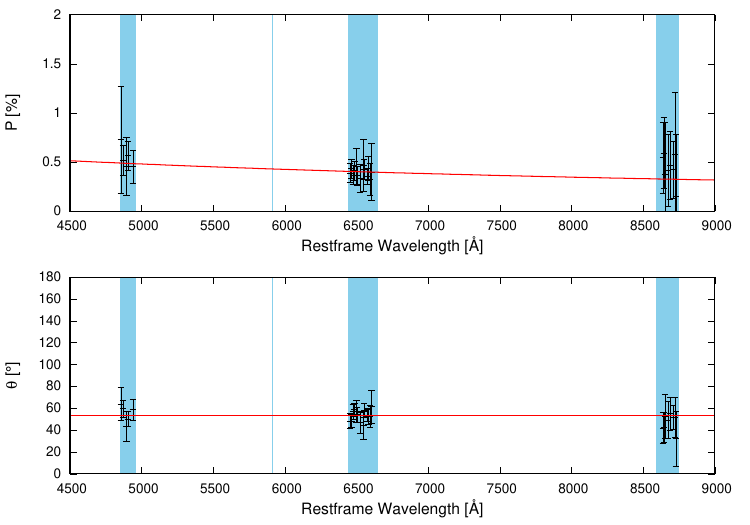}
      \caption{Same as Fig.~\ref{fig:app_ISP_17ahn}, but for SN~2012dh.
    }
    \label{fig:app_ISP_12dh}
\end{figure*}

\begin{figure*}
  \includegraphics[width=2\columnwidth]{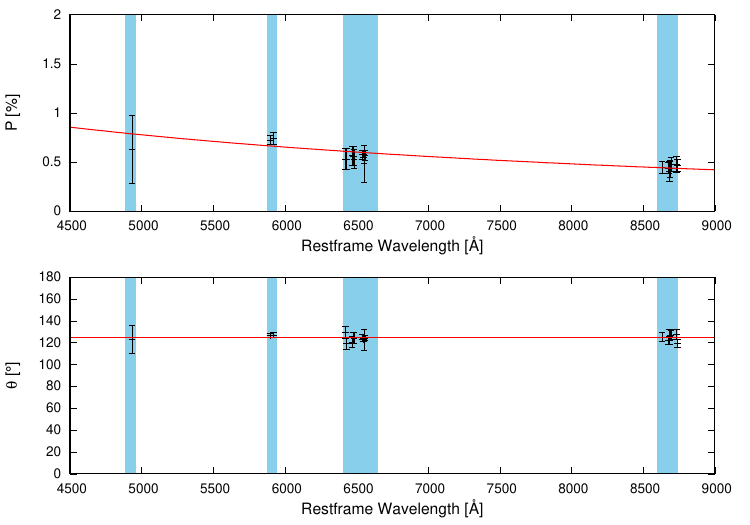}
      \caption{Same as Fig.~\ref{fig:app_ISP_17ahn}, but for SN~2012aw.
    }
    \label{fig:app_ISP_12aw}
\end{figure*}

\begin{figure*}
  \includegraphics[width=2\columnwidth]{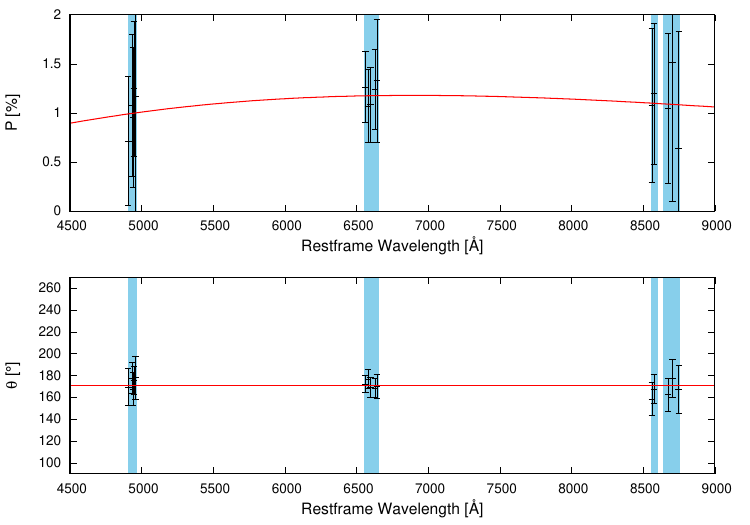}
      \caption{Same as Fig.~\ref{fig:app_ISP_17ahn}, but for SN~2010co.
    }
    \label{fig:app_ISP_10co}
\end{figure*}

\begin{figure*}
  \includegraphics[width=2\columnwidth]{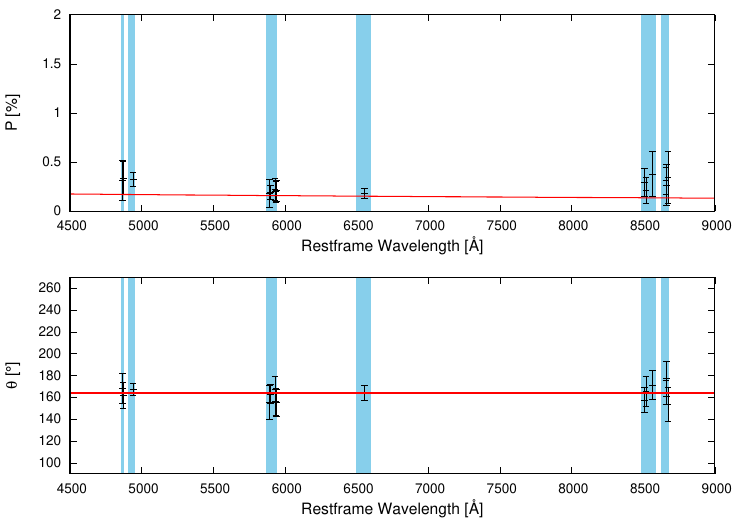}
      \caption{Same as Fig.~\ref{fig:app_ISP_17ahn}, but for SN~2008bk.
    }
    \label{fig:app_ISP_08bk}
\end{figure*}

\begin{figure*}
  \includegraphics[width=2\columnwidth]{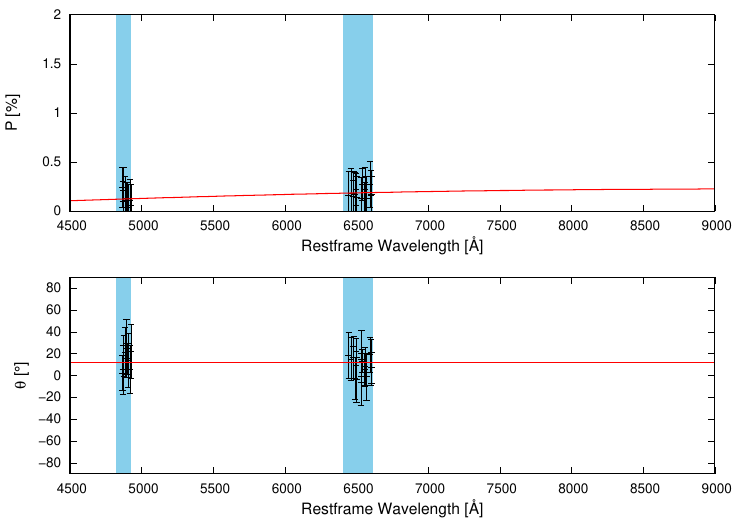}
      \caption{Same as Fig.~\ref{fig:app_ISP_17ahn}, but for SN~2001du.
    }
    \label{fig:app_ISP_01du}
\end{figure*}

\begin{figure*}
  \includegraphics[width=2\columnwidth]{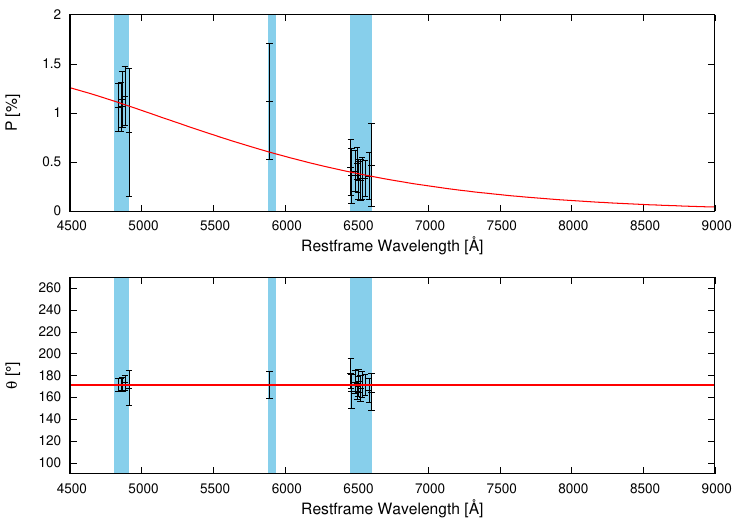}
      \caption{Same as Fig.~\ref{fig:app_ISP_17ahn}, but for SN~2001dh.
    }
    \label{fig:app_ISP_01dh}
\end{figure*}

\end{document}